\documentclass[aps,a4paper,twocolumn,nofootinbib,superscriptaddress]{revtex4}
\usepackage[T1]{fontenc}
\usepackage[latin1]{inputenc}
\usepackage{amsmath,amssymb,color,dsfont}
\usepackage{bm}
\usepackage{bbm}
\usepackage{graphicx}
\usepackage{epsfig}
\usepackage{pdfpages}
\usepackage{color}

\newcommand{\beq}{\begin{equation}}
\newcommand{\eq}{\end{equation}}
\newcommand{\bea}{\begin{eqnarray}}
\newcommand{\ea}{\end{eqnarray}}
\newcommand{\dg}{\dagger}
\newcommand{\la}{\langle}
\newcommand{\ra}{\rangle}
\newcommand{\ov}{\overline}
\newcommand{\nn}{\nonumber}
\newcommand{\partiel}[2]{\frac{\partial #1}{\partial #2}}
\newcommand{\Eq}[1]{Eq.~(\ref{#1})}

\begin{document}

\title{Thermalization of Isolated Bose-Einstein Condensates by 
Dynamical Heat Bath Generation}

\author{Anna Posazhennikova}
\email[Email: ]{anna.posazhennikova@rhul.ac.uk}

\affiliation{Department of Physics, Royal Holloway, University of London,
  Egham, Surrey TW20 0EX, United Kingdom}

\author{Mauricio Trujillo-Martinez}
\affiliation{Physikalisches Institut and Bethe Center for Theoretical Physics,
  Universit\"at Bonn, Nussallee 12, D-53115 Bonn, 
Germany}

\author{Johann Kroha}
\email[Email: ]{kroha@th.physik.uni-bonn.de}

\affiliation{Physikalisches Institut and Bethe Center for Theoretical Physics,
  Universit\"at Bonn, Nussallee 12, D-53115 Bonn, 
Germany} 

\affiliation{Center for Correlated Matter, Zhejiang University, 
Hangzhou, Zhejiang 310058, China} 

\date{\today} 

\begin{abstract}
If and how an isolated quantum system thermalizes despite its unitary time
evolution is a long-standing, open problem of many-body physics. The eigenstate
thermalization hypothesis (ETH) postulates that thermalization happens at
the level of individual eigenstates of a system's Hamiltonian. However, 
the ETH requires stringent conditions to be validated, and it does not 
address how the thermal state is reached dynamically from an initial 
non-equilibrium state.
We consider a Bose-Einstein condensate (BEC) trapped in a double-well potential with an initial population imbalance. We
find that the system thermalizes although the initial conditions violate the ETH requirements. We identify three dynamical regimes.
After an initial regime of undamped Josephson oscillations, 
the subsystem of incoherent excitations or quasiparticles (QP) becomes 
strongly coupled to the BEC subsystem by means of a dynamically generated,
parametric resonance. 
When the energy stored in the QP system reaches its maximum, the number of QPs 
becomes effectively constant, and the system enters a quasi-hydrodynamic
regime where the two subsystems are weakly coupled. In this final regime
the BEC acts as a grand-canonical heat reservoir for the QP system (and vice
versa), resulting in thermalization. We term this mechanism 
dynamical bath generation (DBG). 
\end{abstract}

\maketitle

\section{Introduction}
\label{sec:intro}

Isolated quantum systems pose a challenging problem of quantum physics due to
the unclear mechanism of how these systems reach thermal behavior, as was
experimentally observed
\cite{Trotzky2012,Gring2012,Polkovnikov2011,Yukalov2011}. The experimental
results contradict the common knowledge that unitary time evolution of an
initial pure state,  $|\Psi(t)\rangle=e^{-iHt}|\Psi(0)\rangle$, prohibits
entropy maximization, and as a consequence thermalizaiton should not take
place. A number of quantum thermalization scenarios have been put forward.

One of the most prominent conjectures is the 
eigenstate thermalization hypothesis (ETH) which suggests
that thermalization happens at the level of individual eigenstates
\cite{Deutsch1991,Srednicki1994}. The ETH became very popular after its 
numerical verification for hard-core bosons in two-dimensional lattices 
\cite{Rigol2008,ETH_review}, albeit some systems where it fails have 
been identified \cite{Pozsgay2014}. The ETH is typically restricted to the 
observation of local quantities.

The ETH has been found to be valid even in some integrable systems
\cite{Alba2015}, although thermalization is known in general not to occur
in such cases. The concept of thermalization was adapted to 
systems with non-ergodic dynamics (e.g.  integrable systems), 
by generalized Gibbs ensembles imposing multiple conservation laws 
on average \cite{ETH_review}. A separate branch of research has evolved 
around prethermalization dynamics 
\cite{Kollath2007,Kehrein2008,Kollar2011,Joerg_review} which occur in 
nearly integrable systems with small, integrability-breaking perturbations.

We pursue a different, more generally applicable route to thermalization.
If an isolated quantum system is sufficiently complex, more precisely, if the 
many-body Hilbert space dimension is sufficiently high,
then it is not possible by any experiment
to determine all quantum numbers of a state. The set of 
measured quantum numbers defines a subspace of the total many-body 
Hilbert space. This subspace will be called {\it subsystem}, while the 
remaining subspace of undetermined quantum numbers will serve as a 
{\it thermal bath} or {\it reservoir}. 
The subsystem is then described by a reduced density matrix with 
the reservoir (undetermined) quantum numbers traced out. 
This reduced density matrix will correspond to a statistically mixed 
state, since the system Hilbert space and the reservoir Hilbert space 
are in general entangled. This situation is identical to the canonical
or grand canonical ensemble of an open subsystem coupled to the reservoir. 
In fact, it was shown that any such subsystem of the total 
system is described by the canonical thermal ensemble for the 
overwhelming majority of pure states of the total  
system \cite{Goldstein2006,Reimann2015}.
Thus, according to the second law of thermodynamics and
in spite of the unitary time evolution of the total system,
the subsystem will evolve for long times to the density matrix of a 
(grand) canonical ensemble in thermodynamic equilibrium.   

In the present article we not only study a thermalized state of a subsystem 
in the long-time limit, but we review how such a thermal
state is reached dynamically. We show that the thermalization 
dynamics mentioned above is quite general, if only the Hilbert space 
dimension, i.e., the particle number, is large enough. The coupling between 
bath and subsystem need not be weak, and it is not necessary to define 
separate energy eigenstates of the subsystem and of the bath \cite{Goldstein2006}. 
No restrictions on the initial state (like narrow energy distribution) apply.
Most importantly, it is even valid in cases where either the bath or the 
subsystem Hilbert space is initially not populated, i.e.,
the bath is dynamically generated by the total system's time evolution,
possibly involving multiple time scales \cite{Mauro2009,Mauro2015,Anna2016}. 
We thus term this thermalization dynamics ``dynamical bath generation'' (DBG). 
The DBG mechanism can be understood also as a setup where the subsystem-bath 
coupling evolves in time. 
Initially the coupling constant is zero (the "bath" is absent), whereas during
the bath-generation process the coupling constant reaches its maximum and 
subsequently decays to small, constant values. It is in this final regime 
when one can refer to the total system as being separated into two subsystems, 
each serving as a heat reservoir for the other.

\begin{figure}[!bt]
\begin{center}
\includegraphics[width=0.9\linewidth]{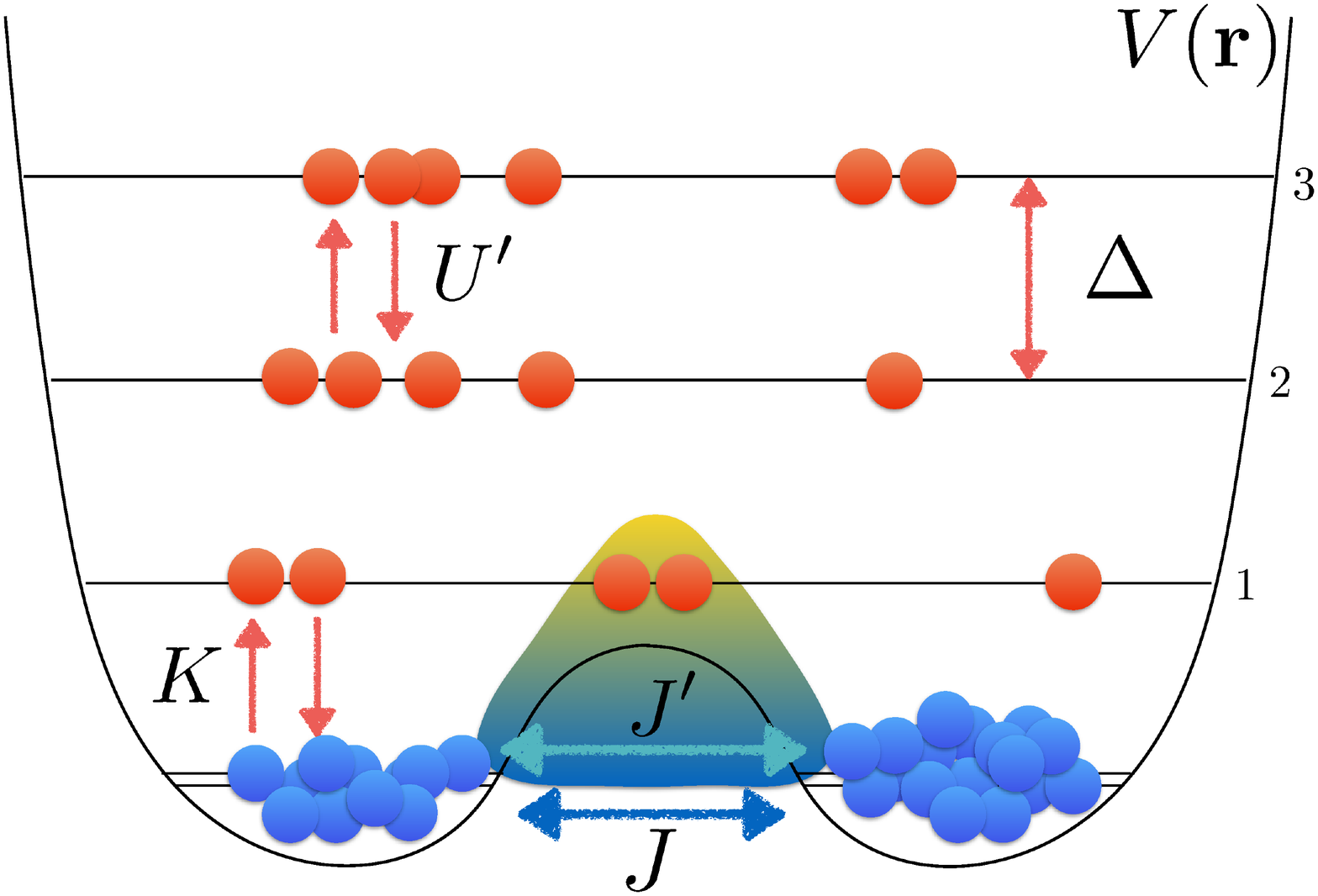}\vspace*{-1.2em} 
\end{center}
\caption{Schematic view of a condensate in a double well potential $V(\mathbf{r})$. The blue dots represent 
atoms in the condensate, while the red dots depict incoherent excitations (quasiparticles) 
out of the condensate. The figure visualizes the energy spacing of trap levels $\Delta$, the bare Josephson coupling $J$
as well as the quasiparticle-assisted Josephson tunnelling $J'$ and the interaction $U$ between particles in 
different levels (see text for more details). Note that the single-particle levels shown are, in general, strongly renormalized by the interactions. }
\label{fig:1}
\end{figure}

\begin{figure}[b]
\begin{center}
\includegraphics[width=0.95\linewidth]{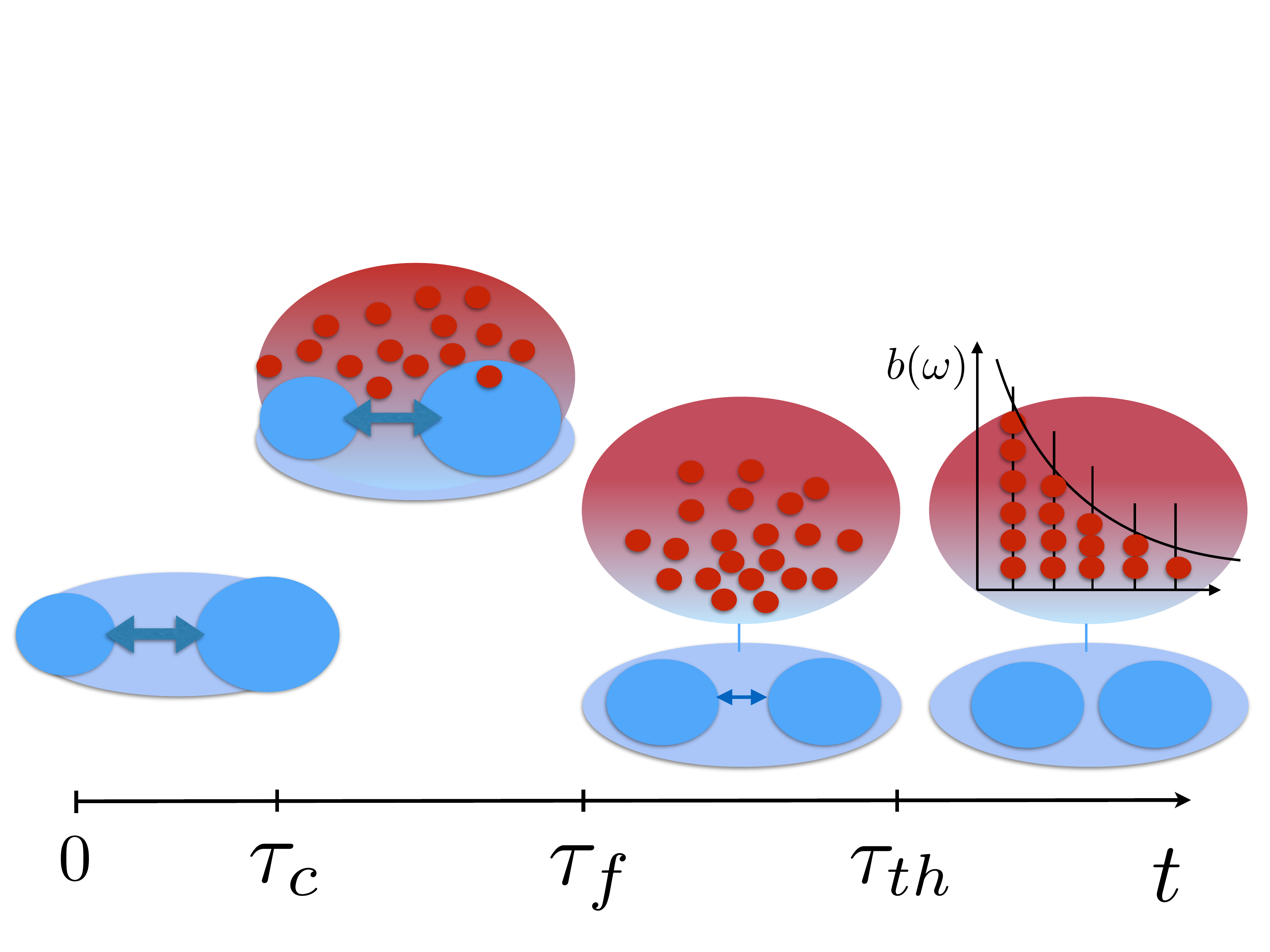}\vspace*{-0.8em} 
\end{center}
\caption{Different time-scales involved in the thermalization of an oscillating Bose gas trapped in a double-well 
potential. $\tau_c$ is the characteristic time scale associated with the creation of incoherent excitations which drastically influence the non-equilibrium BEC dynamics. For $\tau_c<t<\tau_f$ the BEC and QP subsystems are strongly coupled, while for $t>\tau_f$ an effective decoupling takes place imposed by energy conservation. Thermalization occurs with a slow
relaxation rate for $t>t_f$ where the QP subsystem serves as an external bath for the BEC and visa versa. 
The oscillating BECs are depicted in blue, incoherent excitations as red dots.}
\label{fig:2}
\end{figure}

Cold atomic systems are favorable candidates for studying the problem of
closed system thermalization as they can be sufficiently isolated from the
environment and possess an unprecedented degree of tunability. 
They offer the possibility to realize abrupt changes of almost any of the
system parameters (parameter "quenches") thus driving the system 
out of equilibrium in a controlled way. 
As a generic system we consider a Bose-Einstein condensate (BEC) of cold 
atoms trapped in a double-well potential (Bose Josephson junction, 
see Fig.~\ref{fig:1}), 
with initially all atoms in the two single-particle ground states of the 
two wells with a population imbalance $z$. 
We quench the Josephson coupling from $0$ to a finite value $J$ and study 
the resulting dynamics by non-equilibrium quantum field theory methods. 
Interestingly, we identify several time scales which govern the 
non-equilibrium physics, see Fig. \ref{fig:2}. First, 
Josephson oscillations without damping can occur up to a time $t=\tau_c$  
after the quench \cite{Mauro2009,Mauro2015}. 
During a time interval $\tau_c<t<\tau_f$ the condensate (BEC) and the 
quasiparticle (QP) subsystems are strongly coupled via a dynamically 
generated, parametric resonance, indicated by the BEC and the QP spectra 
being strongly correlated with each other \cite{Anna2016}. 
In this regime incoherent excitations are thus created  out of the condensate
in an avalanche-like manner. However, at a freeze-out 
time $\tau_f$ the BEC dynamics effectively decouple from the QP subsystem 
by virtue of total energy conservation, and the BEC and the QP spectra 
become uncorrelated. For $t>\tau_f$ slow, exponential relaxation to 
thermal equilibrium with a relaxation time $\tau_{th}$ occurs due to weak 
coupling of the QP subsystem to the BEC as a grand canonical reservoir,
and vice-versa \cite{Anna2016}.

The article is structured as follows.
In section II we review in some detail the ETH and discuss its 
restrictive assumptions and resulting limitations. Section III contains the
representation of the many-body model Hamiltonian of the Bose gas  
in the basis of trap eigenstates as well as the detailed description 
of the time-dependent Keldysh-Bogoliubov method used to compute the 
dynamics of the coupled system of BEC and incoherent excitations. 
The results are discussed in section IV, describing in detail the 
three different time regimes that are involved in the thermalization 
process of this system. Concluding remarks are given in section V.

\section{Ergodicity and the Eigenstate Thermalization Hypothesis}
\label{sec:ETH}

The ETH provides, within its realm of validity, an explanation why 
isolated quantum systems can behave thermally.  
It also constitutes an attempt at a microscopic, first-principles 
derivation of the ergodic theorem, the basis of equilibrium statistical 
mechanics. Therefore, in this section we briefly recall the ergodic 
theorem and then describe the line of arguments constituting the 
ETH. We also inspect critically the conditions that are necessary 
for this line of arguments to be valid. 

As is well known from  statistical mechanics (see e.g. \cite{Landau}), thermalization of a closed system, isolated from the environment, is rooted in the assumption of ergodicity. The ergodic theorem of classical statistical mechanics states that the statistical or ensemble average $\langle A \rangle$ of a physical observable $A$ is equivalent to 
its long time average $\ov{A(p,q)}$,
\bea
\langle A \rangle&=&
\frac{1}{N!(2\pi\hbar)^{3N}}\int dp dq \rho(p,q)A(p,q)  \nn \\
&=&\lim_{t\rightarrow \infty}\frac{1}{t}\int_0^{t}dt' A(p(t'),q(t')) =\ov{A}\ .
\label{ergodic_class}
\ea
Here, $N$ is the number of particles in the system, $p$ and $q$ are phase space 
coordinates (collectively denoting the coordinates for all $N$ particles), 
and $\rho(p,q)$ is the distribution of the microcanonical ensemble.
For the purpose of proper normalization, the quasiclassical assumption 
has been employed that the particles are indistinguishable and that the 
phase space volume per particle is equal to $(2\pi\hbar)^3$. 
A rigorous derivation of the ergodic assumption Eq. \eqref{ergodic_class} 
has been achieved only in special cases, but a general derivation is still 
lacking \cite{Khinchin,Sinai}. In statistical physics the following heuristic argument is often used \cite{Landau}: Consider a small but still macroscopic subsystem $S_1$ of a given closed system. Let $\Delta p\Delta q$ be a small volume in phase space. Then during a sufficiently long time interval $t$ the subsystem $S_1$ will "visit" $\Delta p\Delta q$ and will spend there some finite time $\Delta t$, so that we can always define a finite probability density
\begin{equation}
\Delta W=\lim_{t\rightarrow \infty}\frac{\Delta t}{t},
\end{equation}
for $S_1$ to be found in the volume  $\Delta p\Delta q$. In this case, it 
is plausible that Eq. \eqref{ergodic_class} will be satisfied with a
corresponding probability distribution of $\Delta W$, see also 
\cite{Birkhoff,Neumann,Singh}. 

\begin{figure}[t]
\begin{center}
\includegraphics[width=0.95\linewidth]{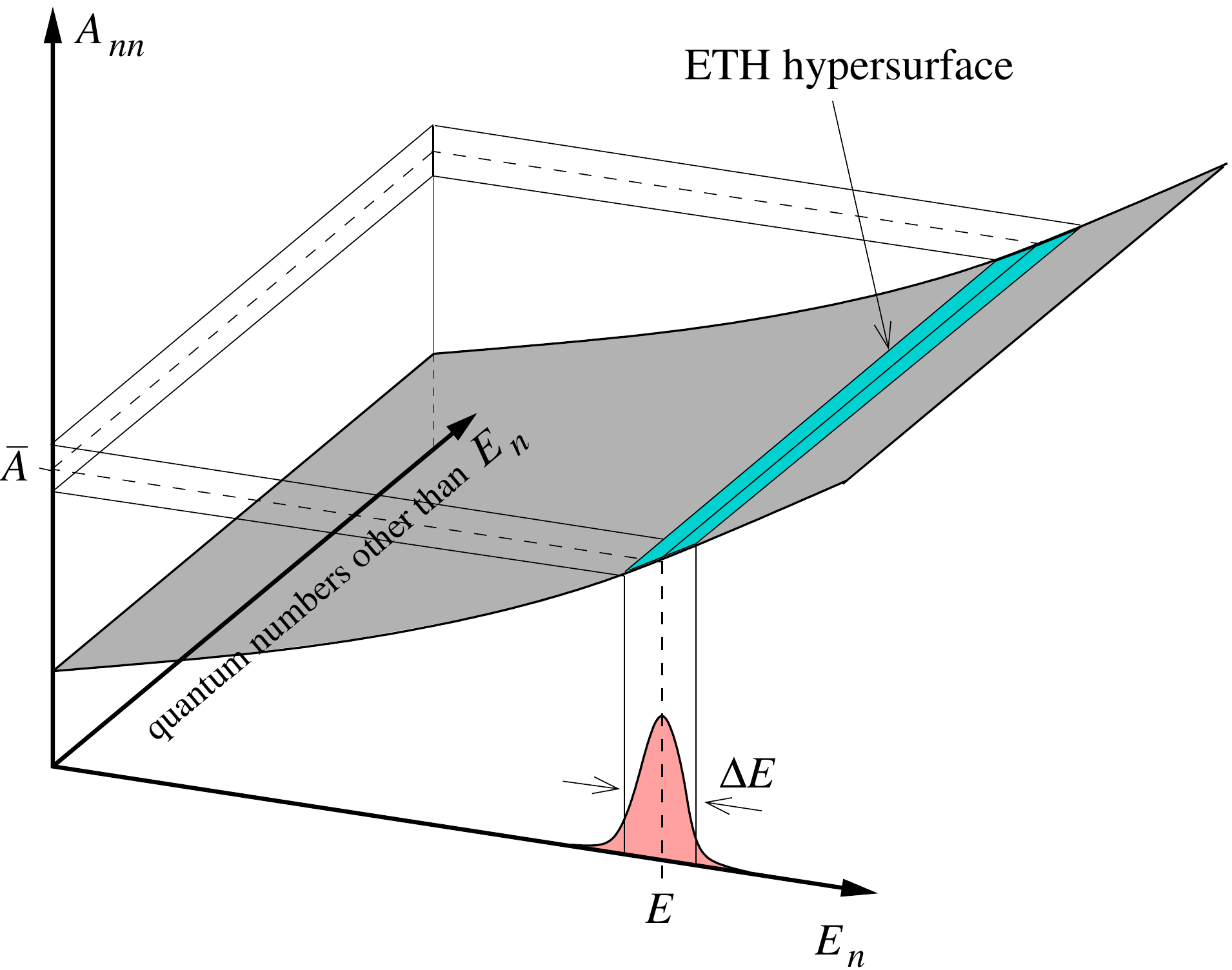}\vspace*{-0.8em} 
\end{center}
\caption{Visualization of the conditions necessary for the ETH to be valid.
In addition to being (essentially) non-degenerate (i), the energy eigenvalues 
$E_n$ must have a narrow spread $\Delta E$ around their mean value (ii).
Furthermore, within the interval $\Delta E$ the $A_{nn}$ must not vary 
strongly (iii), i.e., they should depend continuously on $E_n$, and they
must be independent of other quantum numbers (iv), see text. This defines a 
hypersurface (blue) in the space spanned by all quantum numbers and the 
expectation values $A_{nn}$. It is narrow along the $E_n$ axis and flat 
along all other directions. The ETH scenario applies if the vast majority 
of all $A_{nn}$ lie on this hypersurface.}
\label{fig:3}
\end{figure}

To extend these ideas and concepts to the quantum case, consider now a quantum system described by the Hamiltonian $H$ prepared in an initial state $|\Psi(0)\rangle$. The initial state can be expanded in a complete orthonormal basis $\{|\psi_n\rangle \}$ of eigenstates of the  Hamiltonian,
\bea
H|\psi_n\rangle&=&E_n|\psi_n\rangle, \ \nn \\
|\Psi(0)\rangle&=&\sum_nc_n| \psi_n\rangle,
\ea
with the normalization  $\sum_n|c_n|^2=1$.  
The ETH states that for a physical observable $\hat A$, 
under certain conditions to be discussed below, the long-time avarage of 
the expectation value $\langle\Psi(t)|\hat A|\Psi(t)\rangle$ 
in a many-body state $|\Psi(t)\rangle$ 
is indistinguishable from the thermal average $\langle A\rangle_{mc}(E)$ 
in the microcanonical ensemble with a fixed energy $E$,
\bea
\ov{A}:=\lim_{t\rightarrow \infty}\frac{1}{t}\int_0^{t}dt' 
\langle\Psi(t')|\hat A|\Psi(t')\rangle \stackrel{!}{=} \langle A\rangle_{mc}(E)
\label{time_average} \ .
\ea
The ETH scenario proceeds as follows. For a closed system, the 
unitary time evolution of $|\Psi(t)\rangle$ can be expanded in the basis 
of energy eigenstates, 
\beq
|\Psi(t)\rangle=\sum_n c_n e^{-\frac{i}{\hbar}E_nt}|\psi_n\rangle , 
\label{basis_expansion}
\eq
and the expectation value of $A$ at time $t$ reads, 
\beq
\langle\Psi(t)|\hat A|\Psi(t)\rangle=
\sum_{nm}c_n^*c_me^{-\frac{i}{\hbar}(E_m-E_n)t}A_{nm},
\label{A_expectation} 
\eq
with the matrix elements $A_{nm}=\langle\psi_n|\hat A|\psi_m\rangle$. 
Assuming that (i) the vast majority of the energy eigenvalues are 
non-degenerate, the off-diagonal terms in \Eq{A_expectation} are oscillatory 
and will vanish in the long-time average. One obtains
\beq
\overline{A}=\lim_{t\rightarrow \infty}\frac{1}{t}\int_0^{t}dt' A(t') =\sum_n|c_n|^2A_{nn}.
\label{A_time_average}
\eq
In order to define a microcanonical ensemble with energy $E$ it is now 
necessary to assume that (ii) the distribution of the energy eigenvalues 
in the expansion \Eq{basis_expansion} around the average $E$ is sufficiently
narrow, where the width
\beq
\Delta E=\frac{1}{{\cal N}}\sqrt{\sum_n^{\cal N}(E_n-E)^2} 
\eq
is small on a macroscopic scale, i.e., $\Delta E\ll E$, 
but large enough so that 
there is a large number of energy eigenstates $|\psi_n\rangle$ within 
$\Delta E$.
As two crucial conditions, one furthermore assumes that (iii) the matrix 
elements $A_{nn}$ of the observable $\hat A$ depend continuously on the 
energy eigenvalues $E_n$ and
(iv) they do essentially not depend on any other quantum numbers describing 
the state, see Fig.~\ref{fig:3}. If the conditions (ii), (iii) and (iv) 
are satisfied, then not only the energy eigenvalues $E_n$, but also the 
$A_{nn}$ have a small variation, i.e., they can be assumed constant within the 
set of $|\psi_n\rangle$ contributing to the system's state 
vector $|\Psi(t)\rangle$, 
\beq
A_{nn}\approx A_{E}+\frac{dA_{nn}}{dn} dn= A_{E}+\frac{dA_{nn}}{dn} 
\frac{(E_n-E)}{dE_n/dn}\approx A_E  
\label{A_typical}
\eq
Here, $A_E=\langle\psi_E|\hat A|\psi_E\rangle$ is the matrix element for 
an energy eigenstate with the energy $E$ and $|E_n-E|\leq \Delta E \ll E$.
Thus, the time average from \Eq{A_time_average} can be written 
approximately as 
\beq
\overline{A}\approx\sum_n|c_n|^2A_{E} = A_{E}\ .
\label{A_time_average_ETH}
\eq
Note that the right-hand side of \Eq{A_time_average_ETH}
does not depend on details of the initial conditions, but only on the 
typical energy $E$ of the eigenstates composing $|\Psi(t)\rangle$.  
On the other hand, the microcanonical average of $A$ is 
\beq
\langle A \rangle_{mc}(E)=\frac{1}{\Omega}
\sum_{n:E_n\in[E-\Delta E/2, E+\Delta E/2]} A_{nn} 
\label{micro}
\eq
where $\Omega$ is the number of eigenstates in the narrow energy 
interval $[E,E+\Delta E]$ in the limit $\Delta E\to 0$. Combining 
Eqs.~(\ref{A_typical}), (\ref{A_time_average_ETH}) and (\ref{micro}) 
it follows that the long-time average $\ov{A}$ is equal to  
the quantum mechanical expectation value $A_E$ of one 
representative energy eigenstate and, hence, to the microcanonical average,
\beq
\ov{A} \approx A_E \approx \langle A \rangle_{mc}(E).
\label{quantum_ergodic}
\eq
This is the statement of the ETH.  
It means that the equilibrium thermodynamics 
of an observable $\hat A$ is described by its expectation 
value $A_E$ with respect to a typical energy eigenstate or by its 
long-time average. If Eq.~\eqref{quantum_ergodic} holds, the system 
is called quantum ergodic \cite{Deutsch1991,Rigol2012}. 

However, severe conditions have to be imposed in order 
to reach this conclusion, as seen above:\\[-0.6cm]
\begin{itemize}
\item[(i)] Non-degeneracy of the vast majority of many-body 
eigenenergies $E_n$.\\[-0.62cm]
\item[(ii)] Narrow distribution of the eigenenergies $E_n$ 
around a mean value $E$ on a macroscopic scale:\newline 
$|E_n-E|\lesssim \Delta E<<E$.\\[-0.62cm]
\item[(iii)] Within the width $\Delta E$ all diagonal matrix elements 
of the observable $\hat A$ are approximately equal: 
$A_{nn}\approx A_E$.\\[-0.62cm]
\item[(iv)] These diagonal elements $A_{nn}$ do not depend independently on 
quantum numbers other than the energy eigenvalues $E_n$.\\[-0.6cm]
\end{itemize} 
We now discuss the impact of these assumptions on the applicability of the 
ETH to physical systems. Conditions (i) and (ii) tend to mutually 
exclude each other at first sight. A narrow distribution of the $E_n$ is 
needed in order to define the microcanonical ensemble, but the 
eigenenergies $E_n$, $E_m$ of different states ($n\neq m$) must  
differ sufficiently in order for the offdiagonal terms in \Eq{A_expectation} 
to average out in the long-time average. One expects a relaxation time of 
the order of $1/\Delta E$ which can be macroscopically large. This is 
in contrast to the fast thermalization rates that are usually observed, 
unless conservation laws inhibit thermalization, and that are not 
controlled by $\Delta E$ but by the 
coupling energies of the Hamiltonian (see, e.g., Fig.~\ref{fig:16} and 
Ref.~\cite{Anna2016}). Condition (ii) also restricts the type of initial 
states to which ETH thermalization can apply to those with a narrow 
energy spectrum $\Delta E$. By contrast, many types of initial states,
for instance single-level occupation number eigenstates that appear naturally
as the initial conditions of Josephson trap systems (see below), 
have a broad energy spectrum.
While condition (iii) is plausible for a system 
without a phase transition, condition (iv) clearly imposes a serious 
restriction on the observables that may obey the ETH. It is difficult 
to specify general types of such observables. 

Because of these difficulties in finding general criteria for the applicability 
of the ETH, it has been tested for specific systems using numerical methods, 
such as exact diagonalization \cite{Rigol2008,Roux2009}, time-dependent 
dynamical mean-field theory (tDMFT) \cite{Werner2014,Werner2015}, 
density matrix 
and renormalization group (DMRG) \cite{Schollwoeck2005,Kollath2007}.
In addition, alternative scenarios of thermalization have been put forward, 
see, e.g., \cite{Yukalov2011,Rigol2012} or \cite{Polkovnikov2011} for a review.


\section{Formalism}
\label{sec:formalism}

\subsection{Hamiltonian}
\label{subsec:hamiltonian}

Our goal is to describe Josephson oscillations between weakly-coupled bosonic condensates including effects of quasiparticles. The Josephson effect was originally predicted in superconductors \cite{Josephson1962}, and by now is well studied theoretically \cite{Javanainen1986,Smerzi1997} as well as experimentally \cite{Albiez2005, Levy2007}. Despite a lot of progress there still exist unresolved issues with the experimental results. For example, although in Ref. \cite{Albiez2005} several undamped Josephson oscillations were clearly observed, in other experiments \cite{Thywissen2011} the Josephson particle current was rapidly suppressed. We suggest the quasiparticle damping mechanism to play a crucial role in such a behaviour. Below we present a formalism \cite{Mauro2009,Mauro2015,Anna2016} which, when applied to specific systems, will shed light  on this issue and other problems related to thermalization of isolated closed systems. 

In order to describe a bosonic Josephson junction we start from a weakly interacting Bose gas in a double well potential described by the  well-known  Hamiltonian 
\bea
H=\int d {\bf r} \hat \Psi^{\dg}({\bf r},t)\left(-\frac{\nabla^2}{2m}+V_{ext}({\bf
    r},t)\right)\hat \Psi({\bf r},t) \\ \nonumber
+\frac{g}{2}\int d {\bf r} \hat \Psi^{\dg}({\bf r},t)\hat \Psi^{\dg}({\bf
  r},t)\hat \Psi({\bf r},t)\hat \Psi({\bf r},t),
\label{gen_ham}
\ea
where $\hat \Psi({\bf r},t)$ is a bosonic field operator, and a contact repulsive interaction is implied with the coupling constant $g=4\pi a_s/m$ with $a_s$ being the s-wave scattering length. 
$V_{ext}({\bf r},t)$ is the external double-well trap potential, which in our case is time dependent.
We assume that the barrier between the wells is initially infinitely high, so that the Josephson tunneling between the wells is negligible.  At $t=0$ the barrier  is abruptly lowered down and a sizeable Josephson current will be induced as a result. Such a time dependence of the external potential corresponds to the quenching of the Josephson coupling between the wells, which we can express as
\begin{equation}
J(t)=J\Theta(t),
\label{switch_J}
\end{equation}
where $\Theta(t)$ is the Heaviside step function. The quenching of $J$ results in lowering of the ground state energy of the system by the $\Delta E=J\sqrt{N_1(0)N_2(0)}$, where $N_1(0)$ and $N_2(0)$ are initial occupation numbers of the two wells, which can be quite large. Hence, after the quench two initially separated condensates will be found in an excited state $\Delta E$ above the coupled ground state. We will show that, depending on the system parameters, this energy can suffice to excite quasiparticles out of the BECs  with time-dependent
BEC amplitude playing the role of a perturbation on the
QP system. The QPs will be excited to higher lying {\it discrete} energy levels of the double-well potential, while the two lowest states of the potential are occupied by the BECs.

Before deriving the equations of motion for the field operators $\hat \Psi({\bf r},t)$ and $\hat \Psi({\bf r},t)^{\dagger}$, we expand the operators
in terms of a complete basis  $\mathds{B}=\{\varphi_-,\varphi_+,\varphi_1,\varphi_2, \dots \varphi_M  \}$  of the exact single-particle eigenstates of the double well potential $V_{ext}({\bf r}, t>0)$, i.e. just after the coupling between the wells is turned on. 
Note that the ground state wavefunction has a zero in the barrier between the wells, thus minimizing its energy, i.e., for a symmetric double-well, it
is parity antisymmetric, while the first excited state is symmetric. 
Hence, we denote the ground state wavefunction of the double well by 
$\varphi_-$, the first excited state wavefunction by $\varphi_+$, the 
second excited state by $\varphi_1$ and so on. 
The field operator  in the eigenbasis of the double-well potential is then
\beq
\hat \Psi({\bf r},t)=\phi_1({\bf r})\hat b_{01}(t)+\phi_2({\bf
  r})\hat b_{02}(t)+\sum_{n=1}^{M}\varphi_n({\bf r}) \hat b_n(t)\ ,
\label{field_operator1}
\eq
where we have applied the transformation, 
$\hat b_{01} (t) = (\hat b_- + \hat b_+)/\sqrt{2}$,  
$\hat b_{02} (t) = (\hat b_- - \hat b_+)/\sqrt{2}$ on the operators 
$\hat b_{\pm}(t)$ for particles in the $\varphi _{\pm}$ subspace, 
with the wavefunctions 
$\phi_1({\bf r})=(\varphi_-({\bf r})+\varphi_+({\bf r}))/\sqrt{2}$ and
$\phi_2({\bf r})=(\varphi_-({\bf r})-\varphi_+({\bf r}))/\sqrt{2}$.
Since the $\varphi_+({\bf r})$ ($\varphi_-({\bf r})$) have the same 
(the opposite) sign in the two wells, the $\phi_{1,2}({\bf r})$ are 
localized in the left or right well, respectively, i.e., they 
approximately constitute the ground state wavefunctions of 
the left and right well. We now assume the 
Bogoliubov prescription for the operators describing condensates in each well
\beq
\hat b_{0\alpha}(t) \to a_{\alpha}(t) = 
\sqrt{N_{\alpha}(t)}e^{i\theta_{\alpha}(t)},
\label{cond_wf}
\eq
$\alpha=1,2$, where $N_{\alpha}$ and $\theta_{\alpha}$  are the
number of particles and the phase of the condensate in the left (right) well
of the potential. The field operator finally reads,
\beq
\hat \Psi({\bf r},t)=\phi_1({\bf r}) a_1(t)+\phi_2({\bf
  r}) a_2(t)+\sum_{n=1}^{M}\varphi_n({\bf r}) \hat b_n(t).
\label{field_operator2}
\eq
The Bogoliubov substitution neglects phase 
fluctuations in the ground states of each of the potential wells, while
the full quantum dynamics is taken into account for the excited states, 
$\varphi_n$, $n=1,\ 2,\dots,\ M$ (in the numerical evaluations we will limit the number of levels which
can be occupied by the QPs to $M=5$.). 
This is justified when the BEC particle 
numbers are sufficiently large, $N_{\alpha}\gg 1$, e.g., 
for the experiments \cite{Albiez2005}. When the quantum dynamics due to excitations to upper levels is neglected, only the first two terms from Eq. \eqref{field_operator2} contribute, which is equivalent to a semiclassical two-mode approximation for a condensate in a double well \cite{Milburn1997,Smerzi1997,Gati2007}. The applicability of the semiclassical 
approximation has been discussed in detail in 
Refs.~\cite{Zapata03,Zapata98,Pitaevskii01} and has been tested 
experimentally in Ref.~\cite{Esteve08}. 

We also note that the definition of excited single-particle states is not obvious in the case of a condensate trapped in a double-well  \cite{Smerzi2003,Gati_diss} with the width of the wave-function being in general a function of the number of particles in the well, or total number of particles. Various solutions to this problem and applicability of the approximations are discussed in detail in Ref.  \cite{Ananikian2006}. This issue, however, does not play an important role for the physics we discuss in this work, and we therefore proceed with the expansion \eqref{field_operator2}. 

We can now derive the Hamiltonian of our setup in terms of two condensate amplitudes and quasiparticle operators $\hat b, \hat b^{\dagger}$. For $t>0$ the 
Hamiltonian consists of three contributions
\beq
H=H_{coh}+H_{J}+H_{coll}.
\label{ham}
\eq
$H_{coh}$ includes all local contributions, i.e. all terms
which are bi-linear in the $\hat b_n$-operators and local in 
the well index $\alpha = 1,\ 2$, 
\bea
H_{coh}=\varepsilon_0\sum_{\alpha=1}^2a_{\alpha}^*a_\alpha+\frac{U}{2}\sum_{\alpha=1}^2a_{\alpha}^*a_{\alpha}^*a_\alpha a_\alpha+\sum_{n=1}^M \varepsilon_n\hat b_n^{\dagger} \hat b_n \nn \\
+K\sum_{\alpha=1}^2\sum_{n,m=1}^M \left[a_{\alpha}^*a_{\alpha}\hat b_n^{\dg}\hat b_m+\frac{1}{4}(a_{\alpha}^*a_{\alpha}^*\hat b_n \hat b_m+h.c.)\right],
\ea
where $U$ and $K$ are positive interaction constants, and $\varepsilon_n$ are 
the energies of the $M$ equidistant levels of the double well,
separated by the trap frequency, $\varepsilon_n=n\Delta$. For simplicity we 
neglect hereafter a possible level-dependence of the 
coupling constants. We coin the part of the Hamiltonian $H_{coh}$ "coherent" since it describes only the single-particle dynamics of QPs and therefore  
cannot lead to a decoherence of QPs. 

$H_J$ includes all Josephson-like terms, which are still coherent but are non-local
in the well index, 
\bea
H_{J}&=&-J(a_1^*a_2+a_2^*a_1)+J'\sum_{n,m=1}^M \left[ (a_1^*a_2+a_2^*a_1)\hat b_n^{\dg}\hat b_m \right. \nn \\
&+&\left.\frac{1}{2}(a_1^*a_2^*\hat b_n\hat b_m +h.c.) \right] \ .
\ea
The terms proportional to $J$ are standard Josephson terms also known from the semiclassical approximation \cite{Gati2007}, while  terms proportional to $J'$ describe novel QP-assisted Josephson tunneling events between the wells (see Fig. \ref{fig:1}).

The non-linear collisional terms $H_{coll}$ account for 
full many-body interactions, 
\bea
H_{coll}=\frac{U'}{2}\sum_{n,m=1}^M \sum_{l,s=1}^M \hat b_m^{\dagger}  \hat
b_n^{\dagger} \hat b_l \hat b_s \hspace*{3.3cm}  
\ea
\vspace*{-0.4cm}
\bea 
+R\hspace*{-0.06cm}\left[\sum_{\alpha=1}^2\sum_{n,m,s=1}^M a_{\alpha}^*\hat
  b_n^{\dagger} \hat b_m\hat b_s + \hspace*{-0.06cm} 
  \sum_{\alpha,\beta,\gamma=1}^2\sum_{n=1}^M a^*_{\alpha}a^*_{\beta}a_{\gamma}\hat b_n +h.c.\right] \nn
\label{H_collisions}
\ea

We have introduced the set of parameters in the Hamiltonian \eqref{ham}:
\bea
\varepsilon_0&=&\int d{\bf r}\left[ \frac{\hbar^2}{2m}|\nabla \phi_{1,2}({\bf r})|^2+\phi_{1,2}^2 V_{ext}({\bf r}) \right], \nn \\
U&=&g\int d{\bf r} \, |\phi_{1,2}({\bf r})|^4,  \nn \\
\varepsilon_n&=&\int d{\bf r}\, \varphi_n({\bf r})\left(
-\frac{\nabla^2}{2m}+V_{ext}({\bf r})\right) \varphi_n({\bf r}), \nn \\
U'&=&g\int d{\bf r}\, \varphi_n({\bf r})\varphi_m({\bf r})\varphi_l({\bf r})\varphi_s({\bf r}), \nn \\
J&=&-2\int d {\bf r} \left [ \frac{\hbar^2}{2m}(\nabla \phi_1\nabla \phi_2)+\phi_1 \phi_2 V_{ext}({\bf r}) \right ], \nn\\
K&=&2K_{11nm}=2K_{22nm},  \nn \\
J'&=&2K_{12nm}=2K_{21nm},  \nn \\
R&=&g\int d{\bf r} \phi_{\alpha}({\bf r})\varphi_{n}({\bf r})\varphi_m({\bf r})\varphi_s({\bf r}),
\label{parameters}
\ea 
with $K_{\alpha \beta n m }=g\int d{\bf r}\, \phi_{\alpha}({\bf r})\phi_{\beta}({\bf r})\varphi_n({\bf r})\varphi_m({\bf r})$. 

We will now use the standard non-equilibrium field-theoretical techniques \cite{Rammerbook,Kadanoffbook,Griffinbook} to calculate time-dependence of the following observables: condensate population imbalance
\beq
z(t)=\frac{N_1(t)-N_2(t)}{N_1(t)+N_2(t)},
\eq 
the phase difference between the BECs, $\theta(t)=\theta_2(t)-\theta_1(t)$
and the QP occupation numbers $n_1(t),\ n_2(t),\ \dots, n_M(t)$. 

\subsection{General Quantum Kinetic Equations}
\label{subsec:kin_eq}

The time-dependence of the observables can be calculated from the kinetic equations for the Green's functions of the interacting Bose gas within the Kadanoff-Baym framework \cite{Rammerbook,Kadanoffbook,Griffinbook}. 
As usual, it is convenient to separate the non-equilibrium Green function into its classical and quantum counterparts  and then derive the equations of motion (Dyson equations) for them in the standard way. The classical part ${\bf C}_{\alpha\beta}(t,t')$ is expressed in terms of classical condensate amplitudes $a_{1}(t)$ and $a_{2}(t)$
\beq
{\bf C}_{\alpha\beta}(t,t')=-i\left( \begin{matrix}
a_\alpha(t)a_\beta^*(t') & a_\alpha(t) a_\beta(t') \\
a_\alpha^*(t)a_\beta^*(t')  & a_\alpha^*(t) a_\beta(t')
\end{matrix}
\right),
\label{cond}
\eq
while the quantum part is written in terms of  quasiparticle operators $\hat b_n, \hat b_n^{\dagger}$
\bea
{\bf G}_{nm}(t,t')&=&-i\left( \begin{matrix}
\la T_C\hat b_n(t)\hat b_m^\dg (t') \ra & \la T_C\hat b_n(t)\hat b_m (t') \ra \\ \nn
\la T_C\hat b_n^\dg(t)\hat b_m^\dg (t') \ra  & \la T_C\hat b_n^\dg(t)\hat b_m (t') \ra
\end{matrix}
\right)  \\ 
&=&\left(\begin{matrix} G_{nm}(t,t' ) & F_{nm}(t,t' ) \\ \ov{F}_{nm}(t,t' ) & \ov{G}_{nm}(t,t' )   \end{matrix} \right).
\label{GF_qp}
\ea
Here $\hat T_C$ is time-ordering along the Keldysh contour. Note that at this stage we have already assumed that the position dependence of the Green functions is absorbed in the parameters \eqref{parameters} (for details see \cite{Mauro2015}). 

The general structure of the Dyson equations for these Green's functions is the following
\bea
\int_C d\ov t \left[{\bf G }_0 ^{-1} (t,\ov t) - {\bf S}^{HF} (t,\ov t) \right] {\bf C} (\ov t, t' )  \nn \\
 =  \int_C d\ov t \,{\bf S }(t,\ov t)  {\bf C} (\ov t, t' ),\label{dyson-cond-keldysh} 
\ea
\bea
\int_C d \ov t \left[{\bf G }_0 ^{-1} (t,\ov t)- {\bf \Sigma}^{HF} (t,\ov t) \right] {\bf G} (\ov t, t' )
 = \mathds{1} \delta(t-t') \nn \\ +\int_C d \ov t\, {\bf \Sigma } (t,\ov t)  {\bf G} (\ov t, t' ). 
\label{dyson-noncond-keldysh}
\ea
The term proportional to $\delta(t-t')$ is absent in Eq. \eqref{dyson-cond-keldysh} because of the classical nature of the condensate amplitudes. In the Eqs. \eqref{dyson-cond-keldysh} and \eqref{dyson-noncond-keldysh} 
 we separated the first order in interaction Hartree-Fock self-energies  ${\bf S}^{HF},{\bf \Sigma}^{HF}$ from their second order collisional counterparts ${\bf S},{\bf \Sigma}$. The operator ${\bf G }_0 ^{-1}$ is defined as
\beq
{\bf G }_0 ^{-1}(t,\ov{t})=\delta(t-\ov{t})\left[i\tau_3\frac{\partial}{\partial t}-\left(-\frac{1}{2m}\Delta_1+V_{ext}(t)\right)\mathds{1} \right],
\eq
where 
\beq
\tau_3=\begin{pmatrix} 1 & 0 \\ 0 & -1 \end{pmatrix}.
\eq
All self-energies, including the later appearing ${\boldsymbol{\gamma}},
{\boldsymbol{\Gamma}}$ and ${\boldsymbol{\Pi}}$ have the same $2\times 2$ structure in the Bogoliubov space
\bea
{\bf S}(t,t')=\left(\begin{matrix} S^G(t,t') & S^F(t,t' ) \\ S^{\ov{F}}(t,t' ) & S^{\ov{G}}(t,t' )   \end{matrix} \right),
\ea
where superscripts $G, F, ...$ are references to the corresponding normal and anomalous Green functions in Eq. \eqref{GF_qp}. 

We rewrite the contour integrals in Eqs. \eqref{dyson-cond-keldysh} and \eqref{dyson-noncond-keldysh} as integrals over the real time axis and get the Dyson equation for the condensate Green's function in the form
\bea
 \int\limits_{-\infty}^\infty d \ov{t} \left[{\bf G }_{0,\alpha\gamma} ^{-1} (t,\ov{t}) - {\bf S}^{HF} _{\alpha\gamma}(t,\ov{t}) \right] {\bf C}_{\gamma\beta}(\ov{t} , t'  ) 
= \nn \\  -i \int\limits_{-\infty}^{t} d \ov{t}  {\bm \gamma}_{\alpha\gamma}(t,\ov{t}){\bf C}_{\gamma\beta} (\ov{t} , t'),
\label{dyson_cond}  
\ea
where  ${\bm \gamma}_{\alpha\beta} = {\bm S}^> _{\alpha\beta} - {\bm S}^< _{\alpha\beta}$ (the superscripts "<" and ">" refer to the standard non-equilibrium "lesser" and "greater" self-energies \cite{Rammerbook,Griffinbook}), and  the bare propagator is given by
\beq
{\bf G }_{0,\alpha\beta} ^{-1} (t,t') = \left[i \tau_3 \delta_{\alpha\beta}\partiel{}{t} - \mathds{1} E_{\alpha\beta}\right] \delta(t -t') \ ,
\eq
with $E_{11}=E_{22}=\varepsilon_0$ and $E_{12}=E_{21}=-J$. Hereafter Greek indeces, $\alpha,\,\beta = 1,\,2$, 
refer to the condensates in the left and right wells, and latin indices,
$n,\,m =1,\,2,\,\dots,\,M$, denote the QP levels, we also imply Einstein summation. 

From Eq. \eqref{dyson-noncond-keldysh} we obtain the two equations
\bea
\int\limits_{-\infty}^{\infty} d \ov{t} \left[{\bf G }_{0,nl}  ^{-1} (t,\ov{t}) - {\bf \Sigma}_{nl}^{HF} (t,\ov{t}) \right] {\bf G}_{lm}^\gtrless (\ov{t}, t' )= \nn \\ -i 
\left[\int\limits_{-\infty}^{t_1} d \ov{t}\, {\bf \Gamma}_{nl} (t,\ov{t})  {\bf G}_{lm}^\gtrless (\ov{t}, t' ) - \int\limits_{-\infty}^{t_{1'}}d \ov{t} {\bf \Sigma}_{nl}^\gtrless (t,\ov{t}) {\bf A}_{lm}(\ov{t},t')\right],
\label{dyson-noncond-realtime}
\ea
with 
\beq
{\bf G}_{0,nm}^{-1}(t , t') = \left[i\tau_3 \partiel{}{t} - \varepsilon_{n}\mathds{1} \right]\delta_{nm}\delta(t-t'),
\eq
and
\bea
{\bf G}^<(t,t' )=  -i \left(\begin{matrix} \la \hat b^\dg(t')  \hat b (t ) \ra & \la \hat b(t') \hat b (t ) \ra \\
\la \hat b^\dg (t') \hat b^\dg (t ) \ra &\la \hat b (t') \hat b^\dg (t)  \ra \end{matrix} \right) 
=\nn \\ \left(\begin{matrix} G^<(t,t' ) & F^<(t,t' ) \\ \ov{F}^<(t,t' ) & \ov{G}^<(t,t' )   \end{matrix} \right), \\
{\bf G}^>(t,t' )=-i \left(\begin{matrix} \la \hat b(t)  \hat b^\dg (t' ) \ra & \la \hat b(t) \hat b (t') \ra\\
\la \hat b^\dg (t) \hat b^\dg (t') \ra &\la \hat b^\dg (t) \hat b (t')  \ra \end{matrix} \right) 
=\nn \\ \left(\begin{matrix} G^>(t,t' ) & F^>(t,t' ) \\ \ov{F}^>(t,t' ) & \ov{G}^>(t,t' )   \end{matrix} \right),
\ea 
For practical reasons it is, however, better to work with equations for the spectral function ${{\bf A}_{nm}}=i({\bf G}_{nm}^>-{\bf G}_{nm}^<)$ and the so-called statistical function (see also \cite{Berges_review,Rey2005})
\beq
{{\bf F}_{nm}}=\frac{{\bf G}_{nm}^>+{\bf G}_{nm}^<}{2}=\frac{{\bf G}^K_{nm}}{2},
\eq
here ${\bf G}^K_{nm}$ being the Keldysh component of \eqref{GF_qp}. The
introduction of such symmetrized and antisymmetrized two-point correlators is
not important if we were to reduce the calculation to the first-order
Bogoliubov-Hartree-Fock (BHF) approximation, however, is beneficial for a more general case when second order (in interaction) contributions are taken into account. The derivation then simplifies due to symmetry relations for the propagators and their self-energies, and allows us to rewrite the terms involving higher order processes as "memory integrals".  

In the Bogoliubov space these are $2\times 2$ matrices
\bea
{\bf A}_{nm}(t,t')&=&\left(\begin{array}{cc}
\text{A}^G _{nm}(t,t')& \text{A}^F _{nm}(t,t') \\
\text{A}^{\ov F} _{nm}(t,t') & \text{A}^{\ov G} _{nm}(t,t')
\end{array}
\right), \label{SM:spectral2}\\
{\bf F}_{nm}(t,t')&=&\left(\begin{array}{cc}
\text{F}^G _{nm}(t,t')& \text{F}^F _{nm}(t,t') \\
\text{F}^{\ov F} _{nm}(t,t') & \text{F}^{\ov G} _{nm}(t,t')\end{array}
\right).
\label{statistical2}
\ea
For further derivations we will extensively use the following symmetry relations for the spectral and statistical functions
\bea
\text{A}^{\ov G}(t,t')&=&-\text{A}^{G}(t,t')^*=-\text{A}^{G}(t',t), \nn \\ \nn
\text{A}^{\ov F}(t,t')&=&-\text{A}^{F}(t,t')^*=\text{A}^{F}(t',t)^*, \\ \nn
\text{F}^{\ov G}(t,t')&=&-\text{F}^{G}(t,t')^*=\text{F}^{G}(t',t), \\ 
\text{F}^{\ov F}(t,t')&=&-\text{F}^{F}(t,t')^*=-\text{F}^{F}(t',t)^*.
\label{symmetry}
\ea

The Dyson equations for the spectral and statistical functions are
\bea
\int\limits_{-\infty}^\infty d \ov{t} \left[{\bf G}_{0,n\ell}^{-1}(t , \ov{t}) -{\bf \Sigma}_{n\ell}^{HF} (t , \ov{t}) \right] {\bf A}_{\ell m} (\ov{t} , t') \nn
= \\  -i\int\limits_{t'}^{t} d \ov{t}\, {\bf \Gamma}_{n\ell} (t, \ov{t}) {\bf A}_{\ell m} (\ov{t} , t')
\label{evolution-spectral} \\ \nn
\int\limits_{-\infty}^\infty d \ov{t} \left[{\bf G}_{0,n\ell}^{-1}(t , \ov{t}) -{\bf \Sigma}_{n\ell}^{HF} (t , \ov{t}) \right] {\bf F}_{\ell m} (\ov{t} , t')
= \\ \nn -i\Big{[}\int\limits_{-\infty}^{t} d \ov{t}\, {\bf \Gamma}_{n\ell} (t, \ov{t}) {\bf F}_{\ell m} (\ov{t} , t')\nn \\
\quad\quad-\int\limits_{-\infty}^{t'} d \ov{t}\, {\bf \Pi}_{n\ell}(t, \ov{t}) {\bf A}_{\ell m} (\ov{t} , t')\Big{]},
\label{evolution-statistical}
\ea
where ${\bm \Pi}_{nm} = ({\bm \Sigma}_{nm}^> + {\bm \Sigma}_{nm}^< )/2$ and
${\bf \Gamma}_{nm} = {\bf \Sigma}_{nm}^> - {\bf \Sigma}_{nm}^<$. 
As usual we separated Bogoliubov-Hartree-Fock contributions ${\bf \Sigma}_{nm}^{HF}$ from the second-order contributions describing collisions ${\bf \Sigma}_{nm}$. 

Since the BHF contributions are ${\bf S}^{HF}(t,t')={\bf S}^{HF}(t)\delta(t-t')$, ${\bf \Sigma}^{HF}(t,t')={\bf \Sigma}^{HF}(t)\delta(t-t')$, we can further simplify the Dyson kinetic equations \eqref{dyson_cond}, \eqref{evolution-spectral}, \eqref{evolution-statistical}
\bea
\left[i \tau_3 \delta_{\alpha\gamma}\partiel{}{t} - \mathds{1} E_{\alpha\gamma} - {\bf S}^{HF} _{\alpha\gamma} (t)\right] {\bf C}_{\gamma\beta}(t , t'  ) 
= \nn \\  -i \int\limits_{-\infty}^{t} d \ov{t}  {\bm \gamma}_{\alpha\gamma}(t,\ov{t}){\bf C}_{\gamma\beta} (\ov{t} , t'),
\label{EOMcond}
\ea
\bea
\left[i\tau_3 \delta_{n\ell}\partiel{}{t} - \varepsilon_{n}\delta_{n\ell}\mathds{1}  - {\bf \Sigma}^{HF} _{n\ell}(t)\right]{\bf A}_{\ell m} (t , t')
= \nn \\  -i\int\limits_{t'}^{t} d \ov{t}\, {\bf \Gamma}_{n\ell} (t, \ov{t}) {\bf A}_{\ell m} (\ov{t} , t'),
\label{evolution-spectral1}
\ea
\bea
\left[i\tau_3 \delta_{n\ell}\partiel{}{t} - \varepsilon_{n}\delta_{n\ell}\mathds{1}  - {\bf \Sigma}^{HF} _{n\ell}(t)\right]{\bf F}_{\ell m} (t , t')
=\nn \\  -i\Big{[}\int\limits_{-\infty}^{t} d \ov{t}\, {\bf \Gamma}_{n\ell} (t, \ov{t}) {\bf F}_{\ell m} (\ov{t} , t')\nn \\
\quad\quad-\int\limits_{-\infty}^{t'} d \ov{t}\, {\bf \Pi}_{n\ell}(t, \ov{t}) {\bf A}_{\ell m} (\ov{t} , t')\Big{]}.
\label{evolution-statistical1}
\ea
Eqs.~(\ref{EOMcond}), (\ref{evolution-spectral1}) and
(\ref{evolution-statistical1}) constitute the general equations of motion for
the condensate and the non-condensate (spectral and statistical) propagators.
 They are coupled via the self-energies which are 
functions of these propagators and must be evaluated self-consistently in
order to obtain a conserving approximation.  The higher order interaction terms on the right-hand
side of the equations of motion describe inelastic quasiparticle collisions. They are,
in general, responsible for quasiparticle damping, damping of the
condensate oscillations and for thermalization. We consider them in detail in Section \ref{subsec:collisions}.

\subsection{Bogoliubov-Hartree-Fock Approximation}
\label{subsec:BHF}

\begin{figure}[!bth]
\begin{center}
\includegraphics[width=0.9\linewidth]{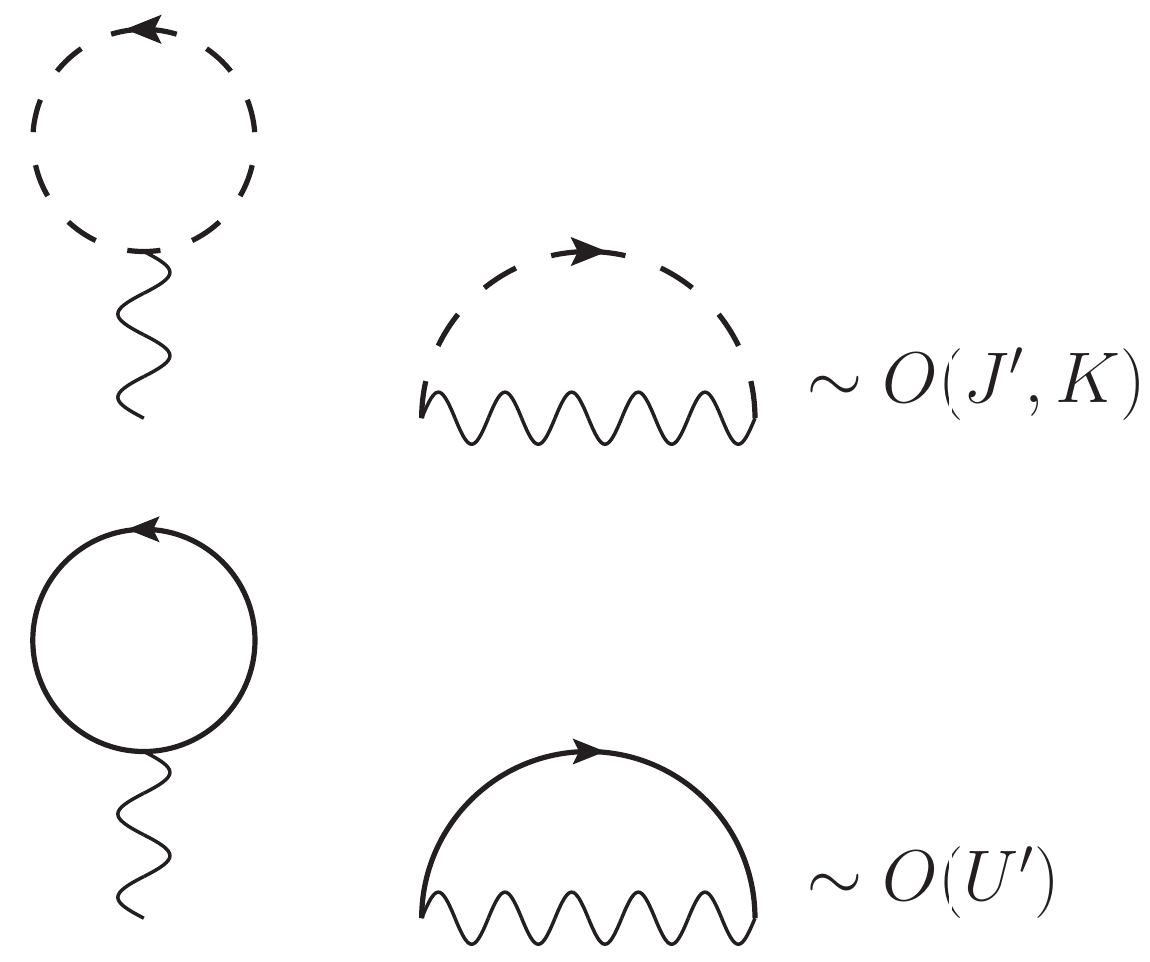}\vspace*{-0.8em} 
\end{center}
\caption{Typical Hartree-Fock diagrammatic contributions to the self-energy  ${\bm\Sigma}^{HF} _{nm}$.
The solid and dashed lines represent  
the $2\times2$ single-particle excitation propagator ${\bf G}$ and the condensate 
propagator ${\bf C}$, respectively. 
The wavy lines denote the interactions $K, J'$ or $U'$, 
depending on  which physical process of the Hamiltonian \eqref{ham} is involved.}
\label{fig:4}
\end{figure}

We now solve Eqs. (\ref{EOMcond}), (\ref{evolution-spectral1}) and
(\ref{evolution-statistical1}) in the first order BHF approximation only. The solutions will  provide us with an interesting initial  insight into the non-equilibrium dynamics of coupled condensates prior to consideration of system's  eventual relaxation to an equilibrium state. The BHF self-energies 
are $2\times 2$ matrices in Bogoliubov space 
\bea
{\bf S}_{\alpha\beta}^{HF}(t)=\left(\begin{matrix} S_{\alpha\beta}^{HF}(t) & W_{\alpha\beta}^{HF}(t) \\
W_{\alpha\beta}^{HF}(t)^* & S_{\alpha\beta}^{HF}(t)^* \end{matrix} \right),
\ea
\bea
{\bf \Sigma}_{nm}^{HF}(t)=\left(\begin{matrix} \Sigma_{nm}^{HF}(t) & \Omega_{nm}^{HF}(t) \\
\Omega_{nm}^{HF}(t)^* & \Sigma_{nm}^{HF}(t)^* \end{matrix} \right).
\ea
${\bf S}^{HF}$ and ${\bf \Sigma}^{HF}$ contain contributions proportional to $U,U', J'$ and $K$ and describe the dynamical shift of the condensate and the
single-particle levels due to  time-dependence of their occupation numbers
and their interactions:
\bea
&&{\bf S}^{HF} _{\alpha\alpha} (t) =\frac{i}{2}U\;\mathrm{Tr}\left[{\bf C}_{\alpha\alpha}(t,t)\right]\mathds{1}+ \nn \\ &&i\frac{K}{2}\sum_{n,m=1}^M \left\{\frac{1}{2} \mathrm{Tr}\left[{\bf F}^< _{nm}(t,t)\right]\mathds{1} + {\bf F}^< _{nm}(t,t)\right\} \nn \\ 
&&{\bf S}^{HF} _{12} (t,t') ={\bf S}^{HF} _{21} (t,t')= \quad \quad \quad \quad \quad \quad \nn \\ &&i\frac{J'}{2}\sum_{n,m=1}^M \left\{\frac{1}{2} \mathrm{Tr}\left[{\bf F}^< _{nm}(t,t)\right]\mathds{1} + {\bf F}^< _{nm}(t,t)\right\}, 
\label{selfenergy-cond-hf}
\ea
and
 \bea
{\bf \Sigma}_{nm}^{HF} (t ,t') = i \frac{K}{2}\sum_{\alpha} \left\{{\bf C}_{\alpha\alpha}(t,t)+\frac{1}{2}\mathrm{Tr}\left[{\bf C}_{\alpha\alpha}(t,t)\right]\mathds{1}\right\} \nn \\
+i\frac{J'}{2}\sum_{\alpha\neq \beta} \left\{{\bf C}_{\alpha\beta}(t,t)+\frac{1}{2}\mathrm{Tr}\left[{\bf C}_{\alpha\beta}(t,t)\right]\mathds{1}\right\}  \nn \\ \nn 
 + i U'\sum_{\ell,s=1}^M \left\{{\bf F}_{\ell s}(t,t)+\frac{1}{2}\mathrm{Tr}\left[{\bf F}_{\ell s}(t,t)\right]\mathds{1}\right\}.\\
\label{selfenergy-non-cond-hf}
\ea
Typical diagrams of the BHF self-energies (e.g. for ${\bf \Sigma}^{HF}$ in Eq. \eqref{selfenergy-non-cond-hf}) are shown in Fig. \ref{fig:4}. 

The equation of motion for the time-dependent condensate amplitude $a_\alpha (t)$ can be obtained by taking the upper left component of Eq. \eqref{EOMcond} and then dividing by $a_\beta ^*(t')$,
\beq
i\partiel{}{t}a_\alpha=\left[ E_{\alpha\gamma}+S^{HF}_{\alpha\gamma} (t) \right]a_\gamma (t)+W^{HF} _{\alpha\gamma}(t)a_\gamma ^*(t). 
\label{eq_cond_HF}
\eq
Equations \eqref{evolution-spectral1} and \eqref{evolution-statistical1} for spectral and statistical propagators decouple in the BHF limit, it is therefore sufficient to consider only Eq. \eqref{evolution-statistical1} in this case
\beq
i\tau_3 \delta_{n\ell}\partiel{}{t}{\bf F}_{\ell m} (t , t') = \left[ \epsilon_{n}\delta_{n\ell}\mathds{1}  + {\bf \Sigma}^{HF} _{n\ell}(t)\right]{\bf F}_{\ell m} (t , t').
\label{evolution-statistical1-hf}
\eq
By taking the difference (or the sum) of Eq. (\ref{evolution-statistical1-hf}) with its hermitian conjugate we obtain two equations
\beq
\begin{split}
& i\left(\partiel{}{t}\text{F}^G _{nm}(t,t') + \partiel{}{t'} \text{F}^G _{nm}(t,t')  \right)
=      \\ & \left[\epsilon_{n}\delta_{n\ell} + \Sigma^{HF} _{n\ell}(t) \right]\text{F}^G _{\ell m}(t,t') -  \Omega^{HF}_{n\ell} (t) \text{F}^F _{\ell m}(t,t')^*- \\
& \text{F}^G _{n\ell}(t,t')\left[\epsilon_{m}\delta_{m\ell} + \Sigma^{HF} _{\ell m}(t') \right] - 
 \text{F}^F _{n\ell}(t,t') \Omega^{HF}_{\ell m} (t') ^* ,
\end{split}
\label{evolution-Fg-hf}
\eq
and
\beq
\begin{split}
& i\left(\partiel{}{t}\text{F}^F _{nm}(t,t') + \partiel{}{t'} \text{F}^F _{nm}(t,t')  \right)
=  \\ & \left[\varepsilon_{n}\delta_{n\ell} + \Sigma^{HF} _{n\ell}(t) \right]\text{F}^F _{\ell m}(t,t') - \Omega^{HF}_{n\ell} (t) \text{F}^G _{\ell m}(t,t')^* + \\
 & \text{F}^F _{n\ell}(t,t')\left[\varepsilon_{m}\delta_{m\ell} + \Sigma^{HF} _{\ell m}(t') \right] + \text{F}^G _{n\ell}(t,t') \Omega^{HF}_{\ell m} (t') ^*.
\end{split}
\label{evolution-Ff-hf}
\eq
The self-energies $\Sigma^{HF}_{nm} (t)$ and $\Omega^{HF}_{nm} (t)$ in these equations are given by
\bea
\Sigma^{HF}_{nm} (t)&=&K(N_1(t)+N_2(t))+J' a_1^*(t)a_2(t) \nn \\
&+& J'a_2^*(t)a_1(t)+2iU'\sum_{s,\ell} \text{F}^G _{s\ell}(t,t),  \\
\Omega^{HF}_{nm} (t)&=&\frac{K}{2}\sum_{\alpha=1}^2a_\alpha(t) a_\alpha(t) 
+J'a_1(t)a_2(t) \nn \\ &+& iU'\sum_{s,\ell} \text{F}^F _{s\ell}(t,t).
\ea

In order to get the final BHF equations we need to evaluate Eqs. (\ref{evolution-Fg-hf}) and (\ref{evolution-Ff-hf}) at equal times, as a result we obtain 
\beq
\begin{split}
i\partiel{}{t}\text{F}^G _{nm}(t,t)&= \left[\varepsilon_{n}\delta_{n\ell} + \Sigma^{HF} _{n\ell}(t) \right]\text{F}^G _{\ell m}(t,t) \\
 &- \text{F}^G _{n\ell}(t,t)\left[\varepsilon_{m}\delta_{m\ell} + \Sigma^{HF} _{\ell m}(t) \right]  \\
&- \Omega^{HF}_{n\ell} (t) \text{F}^F _{\ell m}(t,t)^* - \text{F}^F _{n\ell}(t,t) \Omega^{HF}_{\ell m} (t) ^*, 
\end{split}
\label{evolution-Fg-hf1}
\eq
\beq
\begin{split}
i\partiel{}{t}\text{F}^F _{nm}(t,t) &=\left[\varepsilon_{n}\delta_{n\ell} + \Sigma^{HF} _{n\ell}(t) \right]\text{F}^F _{\ell m}(t,t) \\
& + \text{F}^F _{n\ell}(t,t)\left[\varepsilon_{m}\delta_{m\ell} + \Sigma^{HF} _{\ell m}(t) \right] \\
 &- \Omega^{HF}_{n\ell} (t) \text{F}^G _{\ell m}(t,t)^* + \text{F}^G _{n\ell}(t,t) \Omega^{HF}_{\ell m} (t) ^* .
\end{split}
\label{evolution-Ff-hf1}
\eq
From Eq. \eqref{eq_cond_HF} we get 
\beq
\begin{split}
i\partiel{}{t}a_1 (t) &= \left[\varepsilon_0 + UN_1(t) + iK \sum_{n,m} \text{F}^G _{nm} (t,t)\right]a_1 (t) \\
&-\left[J -iJ'\sum_{n,m} \text{F}^G _{nm} (t,t)\right]a_2 (t) \\
 &+i\left[\frac{K}{2} a_1 ^* (t) + \frac{J'}{2}a_2^* (t)\right] \sum_{n,m} \text{F}^F _{nm} (t,t) 
\end{split}
\label{hartree-fock-eq-cond}
\eq 
The equation for $a_2 (t)$ is obtained from Eq.\eqref{hartree-fock-eq-cond} by  replacing $a_1$ by $a_2$ and visa versa.
We solve differential Eqs. \eqref{evolution-Fg-hf1}, \eqref{evolution-Ff-hf1}  and \eqref{hartree-fock-eq-cond} numerically for different parameters \eqref{parameters},  and different initial conditions $z(0), \theta(0), N_{\textrm{tot}}=N_1(0)+N_2(0)$.  We limit the 
number of levels which can be occupied by the QPs to $M=5$.

\subsection{Collisions in Self-Consistent Second-Order Approximation}
\label{subsec:collisions}

\begin{figure}[b]
\begin{center}
\includegraphics[width=0.9\linewidth]{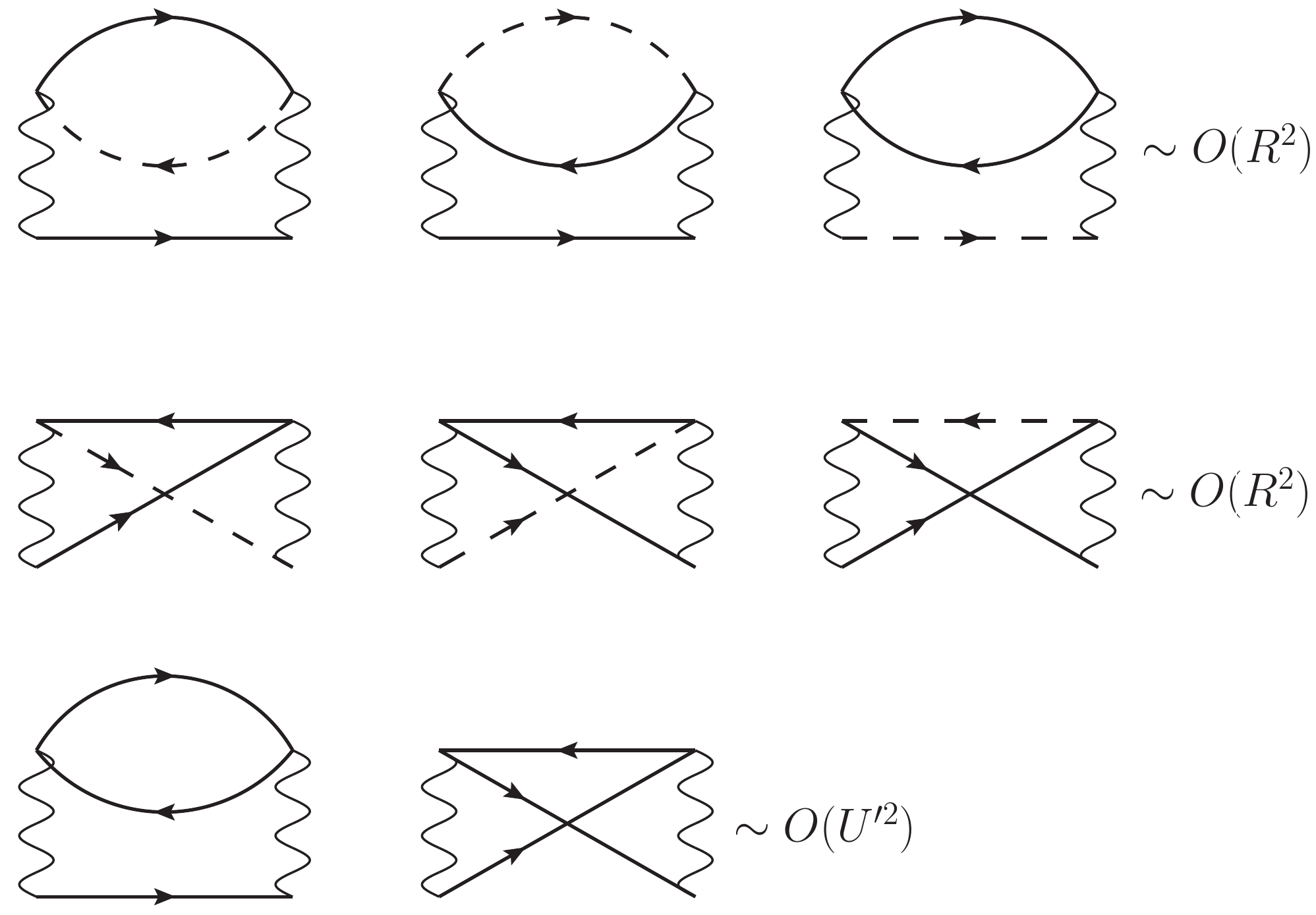}\vspace*{-0.8em} 
\end{center}
\caption{Typical second-order diagrammatic contributions to the QP 
self-energy ${\bm\Sigma}$ \eqref{dyson-noncond-keldysh}.
The solid and dashed lines represent  
the $2\times2$ single-particle excitation propagator ${\bf G}$ and the 
condensate propagator ${\bf C}$, respectively. 
The wavy lines denote the interactions $U,K, J',U'$ or $R$ 
from \eqref{parameters}, depending on  which physical processes of the 
Hamiltonian \eqref{ham} are involved.}
\label{fig:5}
\end{figure}

As we have mentioned,  the BEC oscillations dynamically generate incoherent
excitations (QPs), whose collisions, in turn, may lead to an ultimate thermalization of the system at some finite temperature $T$ controlled by $E_{BEC}(0)$. Although QP generation can be described within the first order BHF approximation, their collisions and eventual equilibration of the system can not. In this section we take into account all second order terms and derive integro-differential equations of motion, which capture the physics of thermalization.  Typical second-order contributions to the quasiparticle self-energies are shown in Fig. \ref{fig:5}. These second-order contributions will lead to a system of IPDE-s, which take into account "memory" effects which are crucial for eventual relaxation of the system. 

Specifically, we need to calculate the non-local self-energies $\gamma, {\bf \Gamma}, {\bf \Pi}$ in the integral parts of Eqs. (\ref{EOMcond}), (\ref{evolution-spectral1}) and
(\ref{evolution-statistical1}). The self-energies are, as usual, matrices in the Bogoliubov space
\bea
{\bf \gamma}_{\alpha\beta}(t,t')&=&\left(\begin{array}{cc}
\gamma^G _{\alpha\beta}(t,t')& \gamma^F _{\alpha\beta}(t,t') \\
\gamma^{\ov F} _{\alpha\beta}(t,t') & \gamma^{\ov G} _{\gamma\beta}(t,t')
\end{array}
\right), \label{gamma} \\
{\bf \Gamma}_{nm}(t,t')&=&\left(\begin{array}{cc}
\Gamma^G _{nm}(t,t')& \Gamma^F _{nm}(t,t') \\
\Gamma^{\ov F} _{nm}(t,t') & \Gamma^{\ov G} _{nm}(t,t')
\end{array}
\right), \label{spectral2}\\
{\bf \Pi}_{nm}(t,t')&=&\left(\begin{array}{cc}
\Pi^G _{nm}(t,t')& \Pi^F _{nm}(t,t') \\
\Pi^{\ov F} _{nm}(t,t') & \Pi^{\ov G} _{nm}(t,t')\end{array}
\right).
\label{SM:statistical2}
\ea

We can now use the symmetry relations \eqref{symmetry} and express the collisional self-energies in terms of only  $\text{A}^G, \text{A}^F$ and $\text{F}^G, \text{F}^F$. We then obtain for $\gamma$ in \eqref{EOMcond}
\bea
\gamma_{\alpha \alpha'}^G(t,t')=R^2\sum_{nls}\sum_{n'l's'}\left(\text{F}_{nn'}^G(t,t')\left\{4\Lambda_{ss'}^{\ell\ell'}[\text{F},\text{F}^*](t,t') \right. \right .\nn \\ \left.  +2\Lambda_{ss'}^{\ell\ell'}[\text{G},\text{G}^*](t,t')\right\} \nn \\ \left.
+\text{A}_{nn'}^G(t,t')\left\{4\Xi_{ss'}^{\ell\ell'}[\text{F},\text{F}^*](t,t')+2\Xi_{ss'}^{\ell\ell'}[\text{G},\text{G}^*](t,t')\right\} \right) \nn \\ 
\gamma_{\alpha \alpha'}^F(t,t')=R^2\sum_{nls}\sum_{n'l's'}\left(\text{F}_{nn'}^F(t,t')\left\{4\Lambda_{ss'}^{\ell\ell'}[\text{G},\text{G}^*](t,t') \right. \right .\nn \\ \left.  +2\Lambda_{ss'}^{\ell\ell'}[\text{F},\text{F}^*](t,t')\right\} \nn \\ \left.
+\text{A}_{nn'}^F(t,t')\left\{4\Xi_{ss'}^{\ell\ell'}[\text{G},\text{G}^*](t,t')+2\Xi_{ss'}^{\ell\ell'}[\text{F},\text{F}^*](t,t')\right\} \right). \nn \\
\ea
In order to make the structure of our equations more transparent, we introduced the shorthand notations
\beq
\begin{split}
\Lambda_{ss'}^{\ell\ell'}[\text{G},\text{F}](t,t')&=\text{A}_{\ell\ell'}^G(t,t')\text{F}^F_{ss'}(t,t') \\ 
&+\text{F}_{\ell\ell'}^G(t,t')\text{A}^F_{ss'}(t,t'),  \\
\Lambda_{ss'}^{\ell\ell'}[\text{G},\text{G}^*](t,t')&=\text{A}_{\ell\ell'}^G(t,t')\text{F}^G_{ss'}(t,t')^*\\ &+\text{F}_{\ell\ell'}^G(t,t')\text{A}^G_{ss'}(t,t')^*,  \\
\Xi_{ss'}^{\ell\ell'}[\text{G},\text{F}](t,t')&=\text{F}_{\ell\ell'}^G(t,t')\text{F}^F_{ss'}(t,t') \\ &-\frac{1}{4}\text{A}_{\ell\ell'}^G(t,t')\text{A}^F_{ss'}(t,t') , \\
\Xi_{ss'}^{\ell\ell'}[\text{G},\text{G}^*](t,t')&=\text{F}_{\ell\ell'}^G(t,t')\text{F}^G_{ss'}(t,t')^* \\ &-\frac{1}{4}\text{A}_{\ell\ell'}^G(t,t')\text{A}^G_{ss'}(t,t')^* , 
\end{split}
\eq
and so on. The remaining components of the self-energy $\gamma$ are related to $\gamma^G$ and $\gamma^F$ by the symmetry relations
\bea
\gamma^G(t,t')^*&=&-\gamma^{\ov G}(t,t')=\gamma^G(t',t), \nn \\
\gamma^F(t,t')^*&=&-\gamma^{\ov F}(t,t')=\gamma^{\ov F}(t',t).
\label{symmetry_gamma}
\ea

For the self-energy ${\bf \Gamma}$ in Eqs. \eqref{evolution-spectral1} and \eqref{evolution-statistical1} we get
\bea
&&\Gamma^G_{nn'}(t,t')=2iR^2\sum_{\alpha\ell s}\sum_{\alpha' \ell' s'}\left( 2a_{\alpha}^*(t) a_{\alpha'}^*(t')\Lambda_{ss'}^{\ell\ell'}[\text{G},\text{F}](t,t') \right. \nn \\
&&+a_{\alpha}^*(t) a_{\alpha'}(t')\Lambda_{ss'}^{\ell\ell'}[\text{G},\text{G}](t,t') \nn \\&&-2a_{\alpha}(t) a_{\alpha'}(t')\Lambda_{ss'}^{\ell\ell'}[\text{F}^*,\text{G}](t,t')  \nn \\ 
&&-\left. 2a_{\alpha}(t) a_{\alpha'}^*(t')\left\{\Lambda_{ss'}^{\ell\ell'}[\text{G},\text{G}^*](t,t')+\Lambda_{ss'}^{\ell\ell'}[\text{F},\text{F}^*](t,t') \right\} \right) \nn \\
&&+(U')^2\sum_{m\ell s}\sum_{m'\ell's'} \left(\text{F}^G_{mm'}(t,t')\left\{4\Lambda_{ss'}^{\ell\ell'}[\text{F},\text{F}^*](t,t') \right. \right. \nn \\ && \left. +2\Lambda_{ss'}^{\ell\ell'}[\text{G},\text{G}^*](t,t') \right\} \nn \\ 
&&+\left.\text{A}^G_{mm'}(t,t')\left\{4\Xi_{ss'}^{\ell\ell'}[\text{F},\text{F}^*](t,t')+2\Xi_{ss'}^{\ell\ell'}[\text{G},\text{G}^*](t,t') \right\} \right) \nn \\
\ea
and
\bea
&&\Gamma_{nn'}^F(t,t')=2iR^2\sum_{\alpha\ell s}\sum_{\alpha' \ell' s'}\left( 2a_{\alpha}^*(t) a_{\alpha'}(t')\Lambda_{ss'}^{\ell\ell'}[\text{G},\text{F}](t,t') \right. \nn \\
&&+a_{\alpha}^*(t) a_{\alpha'}^*(t')\Lambda_{ss'}^{\ell\ell'}[\text{F},\text{F}](t,t') \nn \\&&-2a_{\alpha}(t) a_{\alpha'}^*(t')\Lambda_{ss'}^{\ell\ell'}[\text{G}^*,\text{F}](t,t')  \nn \\ 
&&-\left. 2a_{\alpha}(t) a_{\alpha'}(t')\left\{\Lambda_{ss'}^{\ell\ell'}[\text{G},\text{G}^*](t,t')+\Lambda_{ss'}^{\ell\ell'}[\text{F},\text{F}^*](t,t') \right\} \right) \nn \\
&&+(U')^2\sum_{m\ell s}\sum_{m'\ell's'} \left(\text{F}^F_{mm'}(t,t')\left\{2\Lambda_{ss'}^{\ell\ell'}[\text{F},\text{F}^*](t,t') \right. \right. \nn \\ && \left. +4\Lambda_{ss'}^{\ell\ell'}[\text{G},\text{G}^*](t,t') \right\} \nn \\ 
&&+\left.\text{A}^F_{mm'}(t,t')\left\{2\Xi_{ss'}^{\ell\ell'}[\text{F},\text{F}^*](t,t')+4\Xi_{ss'}^{\ell\ell'}[\text{G},\text{G}^*](t,t') \right\} \right). \nn \\
\ea

While for the self-energy ${\bf \Pi}$ in Eq. \eqref{evolution-statistical1} we have 
\bea
&&\Pi^G_{nn'}(t,t')=2iR^2\sum_{\alpha\ell s}\sum_{\alpha' \ell' s'}\left( 2a_{\alpha}^*(t) a_{\alpha'}^*(t')\Xi_{ss'}^{\ell\ell'}[\text{G},\text{F}](t,t') \right. \nn \\
&&+a_{\alpha}^*(t) a_{\alpha'}(t')\Xi_{ss'}^{\ell\ell'}[\text{G},\text{G}](t,t') \nn \\&&-2a_{\alpha}(t) a_{\alpha'}(t')\Xi_{ss'}^{\ell\ell'}[\text{F}^*,\text{G}](t,t')  \nn \\ 
&&-\left. 2a_{\alpha}(t) a_{\alpha'}^*(t')\left\{\Xi_{ss'}^{\ell\ell'}[\text{G},\text{G}^*](t,t')+\Xi_{ss'}^{\ell\ell'}[\text{F},\text{F}^*](t,t') \right\} \right) \nn \\
&&+(U')^2\sum_{m\ell s}\sum_{m'\ell's'} \left(\text{F}^G_{mm'}(t,t')\left\{4\Xi_{ss'}^{\ell\ell'}[\text{F},\text{F}^*](t,t') \right. \right. \nn \\ && \left. +2\Xi_{ss'}^{\ell\ell'}[\text{G},\text{G}^*](t,t') \right\} \nn \\ 
&&-\left.\frac{1}{2}\text{A}^G_{mm'}(t,t')\left\{2\Lambda_{ss'}^{\ell\ell'}[\text{F},\text{F}^*](t,t')+\Lambda_{ss'}^{\ell\ell'}[\text{G},\text{G}^*](t,t') \right\} \right) \nn \\
\ea
and
\bea
&&\Pi_{nn'}^F(t,t')=2iR^2\sum_{\alpha\ell s}\sum_{\alpha' \ell' s'}\left( 2a_{\alpha}^*(t) a_{\alpha'}(t')\Xi_{ss'}^{\ell\ell'}[\text{G},\text{F}](t,t') \right. \nn \\
&&+a_{\alpha}^*(t) a_{\alpha'}^*(t')\Xi_{ss'}^{\ell\ell'}[\text{F},\text{F}](t,t') \nn \\&&-2a_{\alpha}(t) a_{\alpha'}^*(t')\Xi_{ss'}^{\ell\ell'}[\text{G}^*,\text{F}](t,t')  \nn \\ 
&&-\left. 2a_{\alpha}(t) a_{\alpha'}(t')\left\{\Xi_{ss'}^{\ell\ell'}[\text{G},\text{G}^*](t,t')+\Xi_{ss'}^{\ell\ell'}[\text{F},\text{F}^*](t,t') \right\} \right) \nn \\
&&+(U')^2\sum_{m\ell s}\sum_{m'\ell's'} \left(\text{F}^F_{mm'}(t,t')\left\{2\Xi_{ss'}^{\ell\ell'}[\text{F},\text{F}^*](t,t') \right. \right. \nn \\ && \left. +4\Xi_{ss'}^{\ell\ell'}[\text{G},\text{G}^*](t,t') \right\} \nn \\ 
&&-\left.\frac{1}{2}\text{A}^F_{mm'}(t,t')\left\{\Lambda_{ss'}^{\ell\ell'}[\text{F},\text{F}^*](t,t')+2\Lambda_{ss'}^{\ell\ell'}[\text{G},\text{G}^*](t,t') \right\} \right). \nn \\
\ea

The following symmetry relations apply for the collisional self-energies
\bea
\Gamma^G(t,t')^*&=&-\Gamma^{\ov G}(t,t')=\Gamma^G(t',t), \nn \\
\Gamma^F(t,t')^*&=&-\Gamma^{\ov F}(t,t')=\Gamma^{\ov F}(t',t),  \nn \\
\Pi^G(t,t')^*&=&-\Pi^{\ov G}(t,t')=-\Pi^G(t',t), \nn \\
\Pi^F(t,t')^*&=&-\Pi^{\ov F}(t,t')=-\Pi^{\ov F}(t',t).
\label{symmetry_self}
\ea

The final equations of motion in the collisional case can be then written as follows: 
for the spectral function
\bea
i\frac{\partial}{\partial t}\text{A}^G_{nm}(t,t')=(\varepsilon_n\delta_{n\ell}+\Sigma^{HF}_{n\ell})\text{A}^G_{\ell m}(t,t') \nn \\ -\Omega_{n\ell}^{HF}(t)(\text{A}^F_{\ell m}(t,t'))^* \nn \\
-i\int_{t'}^t d{\ov t}\left[\Gamma^G_{n\ell}(t,{\ov t})\text{A}^G_{\ell m}({\ov t,t'})+\Gamma^F_{n\ell}(t,{\ov t})\text{A}^{\ov F}_{\ell m}({\ov t},t')\right] \nn \\
i\frac{\partial}{\partial t}\text{A}^F_{nm}(t,t')=(\varepsilon_n\delta_{n\ell}+\Sigma^{HF}_{n\ell})\text{A}^F_{\ell m}(t,t') \nn \\ -\Omega_{n\ell}^{HF}(t)(\text{A}^G_{\ell m}(t,t'))^* \nn \\
-i\int_{t'}^t d{\ov t}\left[\Gamma^G_{n\ell}(t,{\ov t})\text{A}^F_{\ell m}({\ov t,t'})+\Gamma^F_{n\ell}(t,{\ov t})\text{A}^{\ov G}_{\ell m}({\ov t},t')\right]. 
\label{spec_fin}
\ea
For the statistical function we get
\bea
i\frac{\partial}{\partial t}\text{F}^G_{nm}(t,t')=(\varepsilon_n\delta_{n\ell}+\Sigma^{HF}_{n\ell})\text{F}^G_{\ell m}(t,t') \nn \\ -\Omega_{n\ell}^{HF}(t)(\text{F}^F_{\ell m}(t,t'))^* \nn \\
-i\int_{0}^t d{\ov t}\left[\Gamma^G_{n\ell}(t,{\ov t})\text{F}^G_{\ell m}({\ov t,t'})+\Gamma^F_{n\ell}(t,{\ov t})\text{F}^{\ov F}_{\ell m}({\ov t},t')\right] \nn \\
+i\int_{0}^t d{\ov t}\left[\Pi^G_{n\ell}(t,{\ov t})\text{A}^G_{\ell m}({\ov t,t'})+\Pi^F_{n\ell}(t,{\ov t})\text{A}^{\ov F}_{\ell m}({\ov t},t')\right] \nn \\
i\frac{\partial}{\partial t}\text{F}^F_{nm}(t,t')=(\varepsilon_n\delta_{n\ell}+\Sigma^{HF}_{n\ell})\text{F}^F_{\ell m}(t,t') \nn \\ -\Omega_{n\ell}^{HF}(t)(\text{F}^G_{\ell m}(t,t'))^* \nn \\
-i\int_{0}^t d{\ov t}\left[\Gamma^G_{n\ell}(t,{\ov t})\text{F}^F_{\ell m}({\ov t,t'})+\Gamma^F_{n\ell}(t,{\ov t})\text{F}^{\ov G}_{\ell m}({\ov t},t')\right] \nn \\
+i\int_{0}^t d{\ov t}\left[\Pi^G_{n\ell}(t,{\ov t})\text{A}^F_{\ell m}({\ov t,t'})+\Pi^F_{n\ell}(t,{\ov t})\text{A}^{\ov G}_{\ell m}({\ov t},t')\right]. 
\label{stat_fin}
\ea
And, finally, for the condensate amplitudes we have 
\bea
i\frac{\partial}{\partial t}a_{\alpha}(t)=-Ja_{\beta\neq\alpha}(t)+S_{\alpha\beta}^{HF}a_{\beta}(t)+W_{\alpha\beta}^{HF}a_{\beta}^*(t) \nn \\
-i\int_{0}^td{\ov t}\left[\gamma_{\alpha\beta}^G(t,{\ov t})a_{\beta}({\ov t})+\gamma^F_{\alpha\beta}(t,{\ov t})a^*_{\beta}({\ov t}) \right],
\label{cond_fin}
\ea
with the self-energies specified at the beginning of the section. We also used the symmetry relations \eqref{symmetry} in order to  express all quantities appearing in the equations of motion with the later time argument on the left side. This implies that we only need to know the solution of the previous steps while evolving the equations in time (this is beneficial for the numerical implementation - see Appendix \ref{numerics}). The integro-differential equations \eqref{spec_fin}, \eqref{stat_fin} and \eqref{cond_fin} are then solved for different values of parameters \eqref{parameters} and different initial values  
$z(0), \theta(0), N_{\textrm{tot}}=N_1(0)+N_2(0)$. 

The kinetic equations \eqref{spec_fin}, \eqref{stat_fin} and \eqref{cond_fin} are derived here for the case of two weakly-coupled trapped condensates, however, since our formalism is quite general, they can be easily extended to the case of several weakly coupled condensates and to optical lattices, where non-equilibrium dynamics in the weakly interacting regime can be analysed in detail \cite{Mauro2017}.


\section{Quasiparticle Creation and Thermalization 
Dynamics}
\label{sec:results}

\subsection{Results within Bogoliubov-Hartree-Fock Approximation}
\label{subsec:BHF_results}

In this section we present numerical solutions of Eqs. \eqref{evolution-Fg-hf1}, \eqref{evolution-Ff-hf1}  and \eqref{hartree-fock-eq-cond}
for
total number of particles $N_{\textrm{tot}}=500000$, level spacing $\Delta$,  
interactions $U,\,U',\,K,\,J'$. As initial conditions 
we take all particles in the condensate, $N_1(0)+N_2(0)=N_{\textrm{tot}}$, 
with population imbalance $z(0)$ and phase difference 
$\theta(0)$ between the BECs in the two wells
at time $t=0$, that is, 
\bea
a_{\alpha}(0)&=&\sqrt{N_{\alpha}(0)}e^{i\theta_{\alpha}(0)}, \qquad \alpha=1,2 \nn \\
\text{F}^G_{nm}(0,0)&=&-\frac{i}{2}\delta_{nm}  \\
\text{F}^F_{nm}(0,0)&=&0. \nn
\ea
Note that the large particle number we chose  is at least two orders of magnitude larger then the number of atoms in realistic experiments \cite{Albiez2005,Thywissen2011}. 
We did it in order to make the distinction between BEC and incoherent excitation especially pronounced.

It is convenient to express all energies in units of the bare 
Josephson coupling $J$: $u=UN_{\textrm{tot}}/J$,
$u'=U'N_{\textrm{tot}}/J$, $k=KN_{\textrm{tot}}/J$, $j'=J'N_{\textrm{tot}}/J$,
$r=RN_{\textrm{tot}}/J$, whereas time $t$ is given in units of $1/J$. 
We emphasize that in general these parameters are determined by the trap and by the scattering length  of the atoms  in the trap (see Eqs. \eqref{parameters}). 
In this work we choose certain characteristic values in order to demonstrate the typical thermalization dynamics in the presence of quasiparticles.
We calculate the time dependence of the condensate population imbalance 
$z(t)$, the phase difference 
$\theta(t)$, the QP occupation numbers $n_m(t)$
in the excited trap states, $m=1,2,.., 5$, and the total occupation number
\beq
n_{\textrm{tot}}(t)=\sum_{m=1}^5n_m(t)=-\sum_{m=1}^5\left[\text{Im}\text{F}_{mm}^G(t,t)-\frac{1}{2} \right].
\label{QP_tot}
\eq
The QP occupation numbers are normalized by the total number of 
particles $N_{\textrm{tot}}$, which is conserved. 

Our main finding in the BHF regime is that there exists a characteristic time
scale $\tau_c$ associated with the creation of incoherent excitations 
(QPs) out of the condensate. When $J'=K=0$, this time scale is infinite, 
and the system performs undamped Josephson oscillations, described by the 
semiclassical two-mode approximation \cite{Smerzi1997}. For non-zero interaction
parameters, however, a qualitatively different dynamics sets in at the 
characteristic time $\tau_c$. At this time, QPs get excited, and the 
dynamics becomes dominated by fast QP Rabi oscillations between the discrete 
trap levels, which in turn drive the BEC oscillations \cite{Mauro2009}. 
We note that after the time $\tau_c$ inelastic QP collisions will become 
important and will be taken into account in section \ref{subsec:thermalization}. The 
collisionless regime that exists up to $\tau_c$ and, therefore, 
the time scale $\tau_c$ itself can be described by the BHF approximation.

\begin{figure}[t]
\begin{center}
\includegraphics[width=0.9\linewidth]{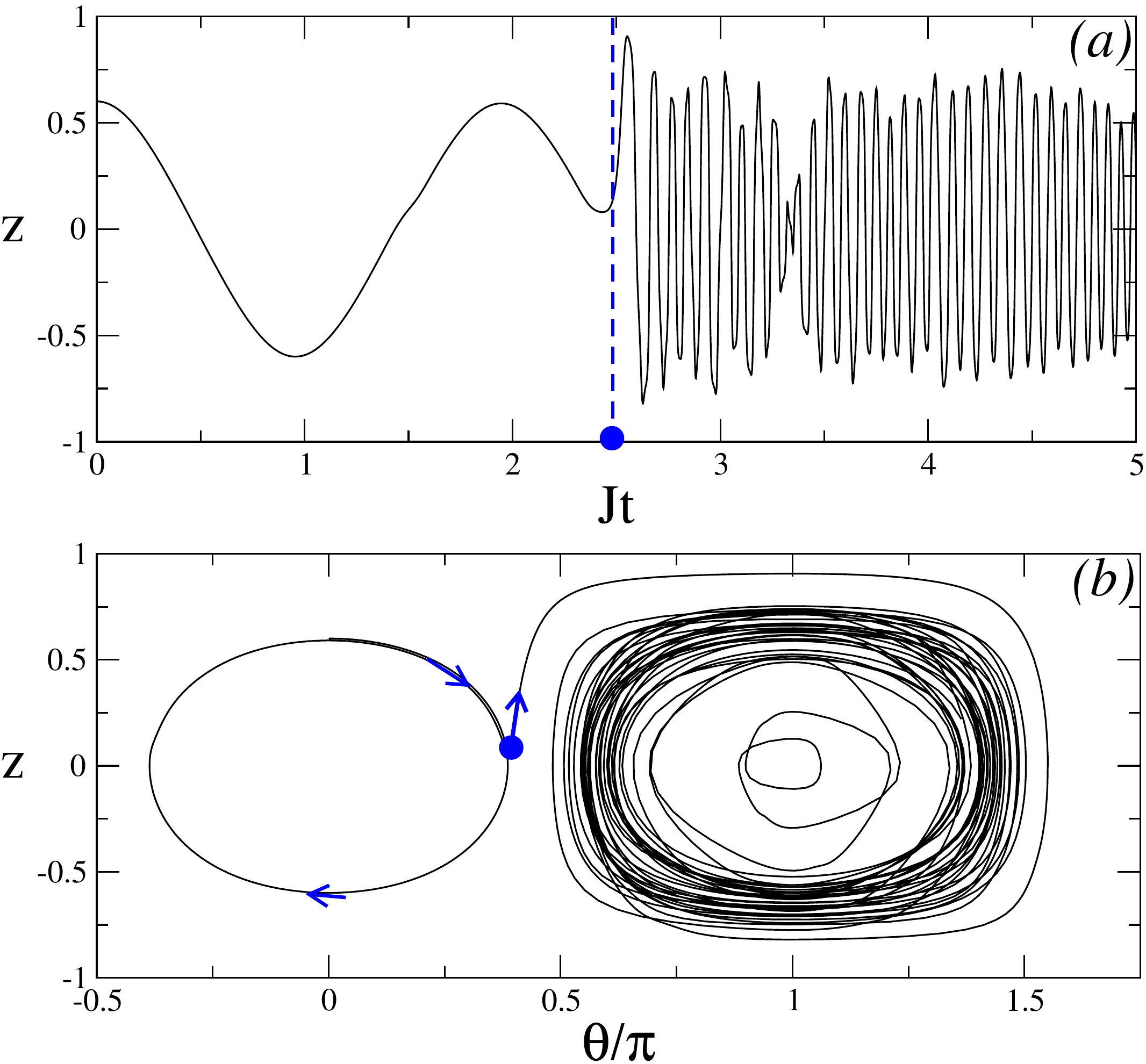}\vspace*{-0.8em} 
\end{center}
\caption{(a) Time-dependent population imbalance $z(t)$ and (b) phase space 
portrait for $\Delta=20, u=u'=5, j'=60, k=0$, and initial conditions 
$z(0)=0.6, \theta(0)=0$. The characteristic time scale $\tau_c$ is marked by
the dashed vertical line, and $\tau_c$ is shown by the thick dot in both
panels. The arrows indicate the clockwise direction of time evolution 
along the phase space trajectory.}
\label{fig:6}
\end{figure} 

\begin{figure}[b]
\begin{center}
\includegraphics[width=0.92\linewidth]{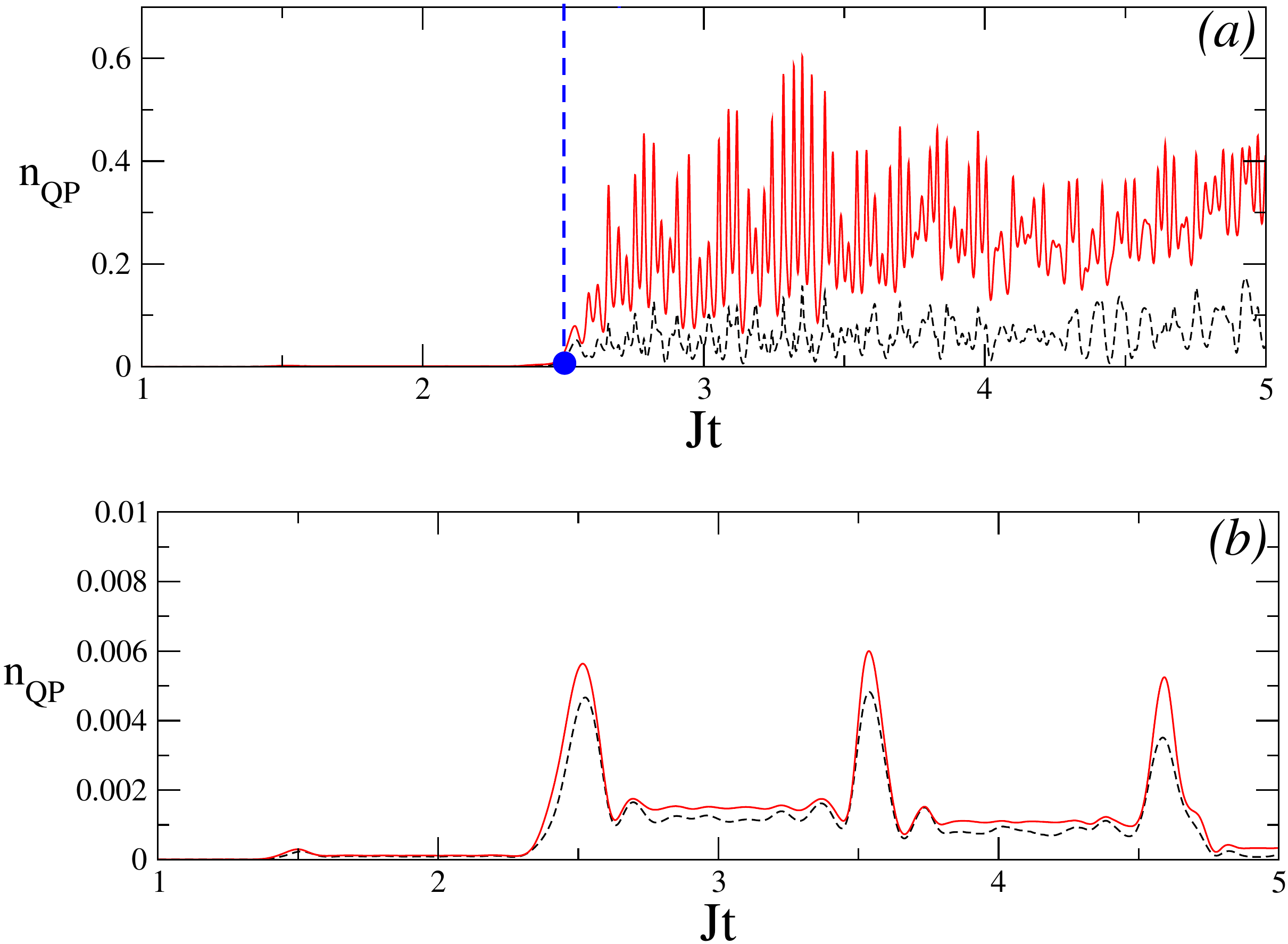}\vspace*{-0.8em} 
\end{center}
\caption{QP occupation numbers $n_{\text{QP}}$ (a) for the same parameters as
  in Fig. \ref{fig:6}, (b) for the parameters as in
  Fig. \ref{fig:8}. The dashed, black lines represent the QP
  occupation number of the first QP level: $n_1(t)$, while the solid, red
  lines correspond to the sum of all five levels $n_{\textrm{tot}}(t)$
  \eqref{QP_tot}. In (a) the dashed vertical line marks the onset of the QP
  dominated regime at $t=\tau_c$. In (b) $\tau_c$ is not identifiable.}
\label{fig:7}
\end{figure} 
  
In Fig. \ref{fig:6} we show how the QP creation sets in for a
certain choice of the parameters in Eq.~\eqref{parameters}. 
In Fig. \ref{fig:6} (a) the commencement of the QP-dominated
dynamics is indicated by the vertical dashed line, and $\tau_c$ by a thick
dot. For $t<\tau_c$ the junction exhibits undamped Josephson oscillations with
a frequency $\omega_J$ which can be estimated from the two-mode approximation
as $\omega_J\approx 2J\sqrt{1+u/2}$ \cite{Smerzi1997}. At $t>\tau_c$ a 
substantial amount of QPs is abruptly created as seen in Fig.~\ref{fig:7}(a), 
and fast Rabi oscillations between the QP levels govern the dynamics.

It seems surprising at first sight that for a discrete 
spectrum $\tau_c$ can be non-zero. 
It means that QPs are not excited immediately, even though the
initial state with $z(0)\neq 0$ is a highly excited state with a macroscopic 
excitation energy $E_{BEC}(0)$, sufficient to excite QPs. $E_{BEC}(0)$ is 
proportional to $z^2(0)N_{\textrm{tot}}J$, as derived in the next section
\ref{subsec:thermalization}. The reason is that the condensate oscillations  
act as a periodic perturbation with frequency $\omega_J$ on the QP subsystem.
Therefore, for $\omega_J<\Delta_{\textrm{eff}}$ (where $\Delta_{\textrm{eff}}$ is the effective 
level spacing, renormalized by all interactions) QPs cannot be excited in low
order time-dependent perturbation theory. QP excitations are possible 
only in higher orders which is a highly non-linear process and leads to the 
abrupt creation of QPs at $t=\tau_c$. For $t>\tau_c$, QP collisions are 
expected to ultimately thermalize the system, as described in 
section \ref{subsec:thermalization}.

\begin{figure}[t]
\begin{center}
\vspace*{-0.3cm}
\includegraphics[width=0.9\linewidth]{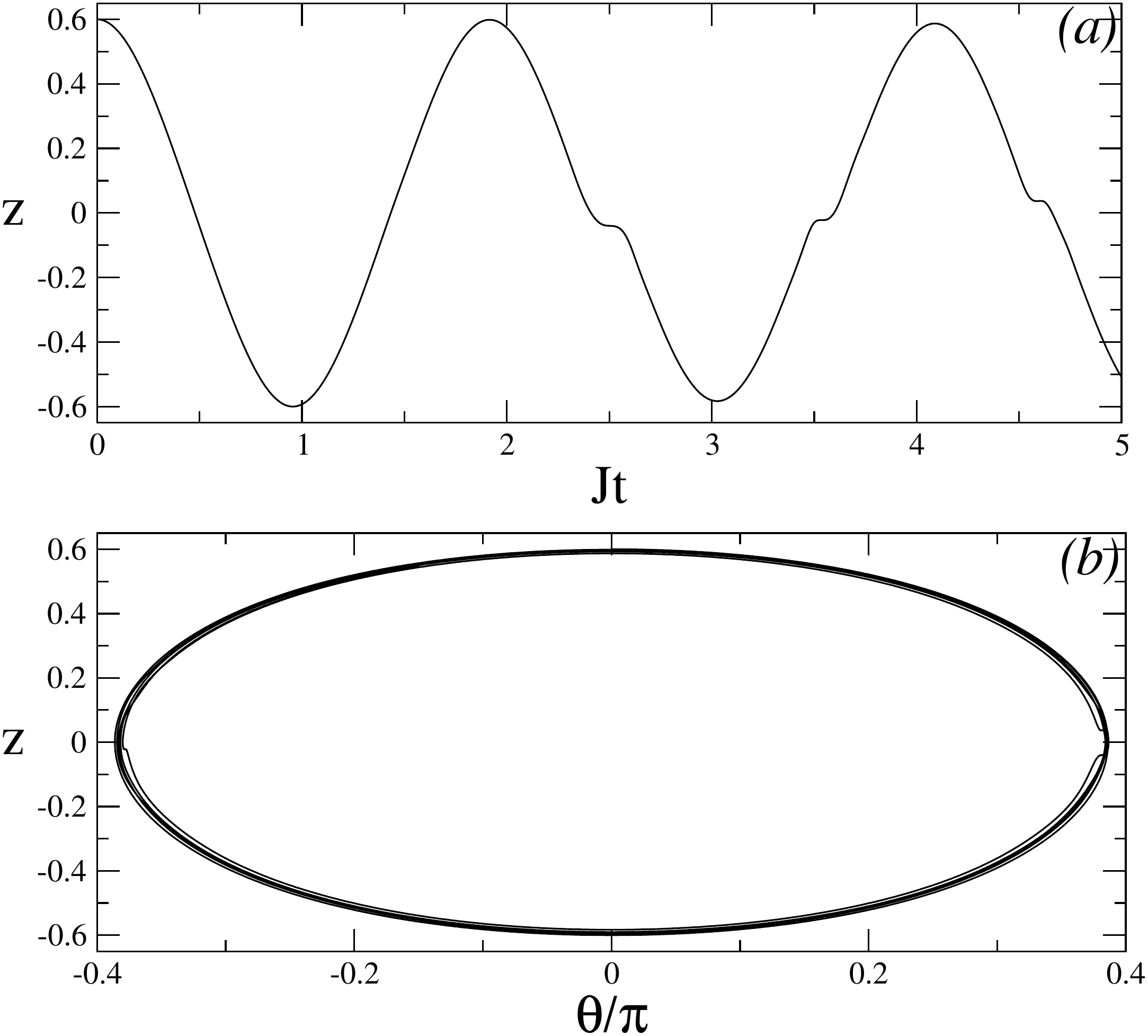}\vspace*{-0.8em} 
\end{center}
\caption{(a) Time-dependent population imbalance $z(t)$ and (b) phase space
  portrait  for $\Delta=20, u=u'=5, j'=60, k=0.38$, and initial conditions
  $z(0)=0.6, \theta(0)=0$. $\tau_c$ is beyond the observation time in this case. }
\label{fig:8}
\end{figure}

\begin{figure}[b]
\vspace*{+1.8em}
\begin{center}
\includegraphics[width=0.95\linewidth]{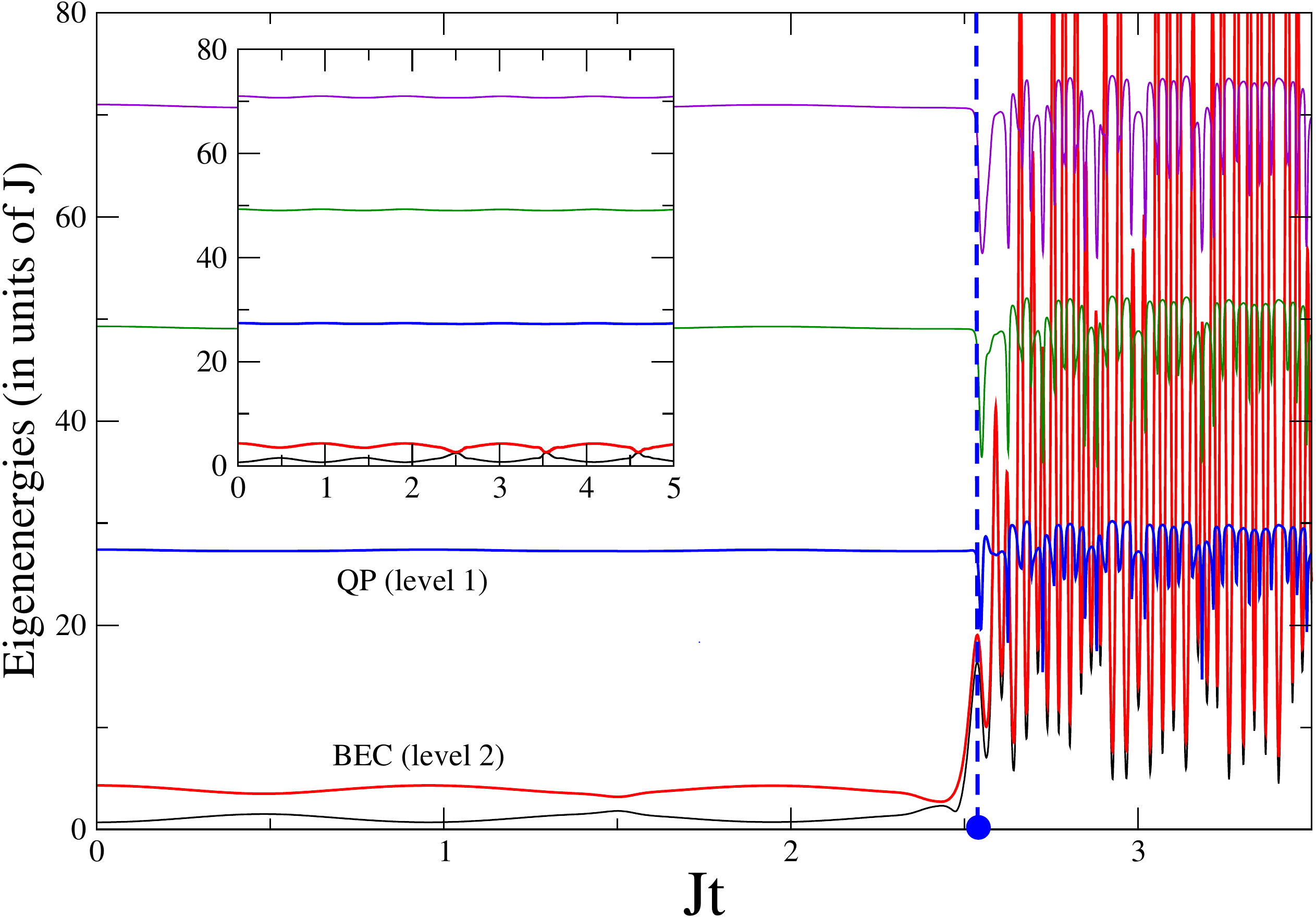}\vspace*{-0.8em} 
\end{center}
\caption{Renormalized single-particle levels of Hamiltonian \eqref{ham} in BHF
  approximation for the parameters of
  Fig. \ref{fig:6}:  $\Delta=20, u=u'=5, j'=60, k=0$. The onset of
  the QP-dominated regime is indicated by the dashed, vertical line, and $\tau_c$
  by the thick dot. The two lowest levels are the condensate levels (black
  and red), the first QP levels is marked as "level 1" (in blue).  In the
  inset the instantaneous eigenenergies for the parameter set of 
  Fig.~\ref{fig:8} are shown. $\tau_c$ is unobservably large 
  in this case.}
\label{fig:9}
\end{figure} 

Another remarkable phenomenon associated with the QP dynamics is that at 
$\tau_c$ that Josephson junction undergoes a $0-\pi$ transition, as can be 
seen from the phase space portrait in Fig. \ref{fig:6}(b). Prior to
$\tau_c$, the phase difference oscillates
around $\theta(0)=0$, whereas for $t>\tau_c$ it oscillates around $\pi$. This
behaviour can be understood recalling the analogy to a driven
oscillator. While for $t<\tau_c$ the Josephson junction oscillates at its
natural frequency $\omega_J$, for $t>\tau_c$ it is driven by the QP Rabi 
oscillations with frequencies $\omega_R\approx \Delta_{\textrm{eff}}>>\omega_J$ far 
above its resonance frequency and, thus, has a phase shift of $\pi$ 
with respect to the QP density as a driving force. The $0-\pi$ transition 
should be detectable in phase sensitive experiments \cite{Albiez2005}
when QPs are excited.

The time  scale $\tau_c$ depends strongly on the parameters of
Eq.~\eqref{parameters}. Consequently, it is sensitive to details of the
experimental setup, which can be realized in very dissimilar ways
\cite{Albiez2005,Thywissen2011,Lappe2017}.  In Fig. \ref{fig:8} we
have chosen a different value of the BEC-QP coupling $k$, which results in a
drastic suppression of QPs and therefore their negligible effect on the
junction dynamics. In this case the Bose Josephson junction is well described
within the semiclassical two-mode approximation \cite{Smerzi1997}, although a
small density of incoherent excitations may be excited intermediately. 
Such a low QP density decays again, as shown in Fig. \ref{fig:7}(b), and
is not sufficient to induce fast Rabi oscillations or a $0-\pi$ transition,
see~Fig\ref{fig:8}(b).

\begin{figure}[t]
\vspace*{-1.0em}
\begin{center}
\includegraphics[width=0.9\linewidth]{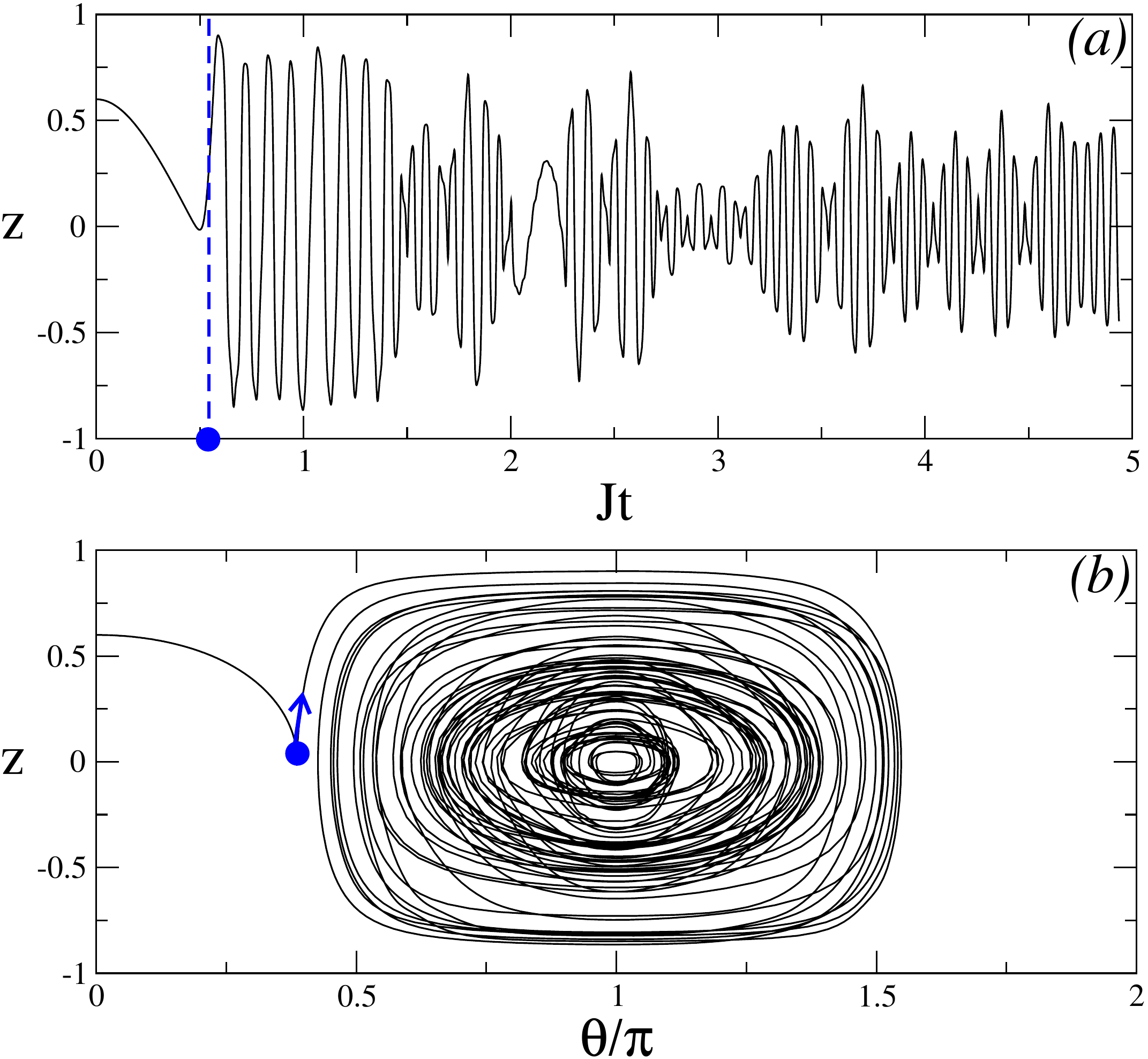}\vspace*{-0.8em} 
\end{center}
\caption{(a) Time-dependent population imbalance $z(t)$ and (b) phase space
  portrait for $\Delta=15, u=u'=5, j'=60, k=0$, and initial conditions
  $z(0)=0.6, \theta(0)=0$. The characteristic time scale $\tau_c$ is marked by
  the dashed vertical line, and $\tau_c$ is shown by the thick dot in both
  figures. The arrow indicates the direction of the time evolution along the 
  phase space tranjectory.}
\label{fig:10}
\end{figure} 

\begin{figure}[b]
\begin{center}
\includegraphics[width=0.95\linewidth]{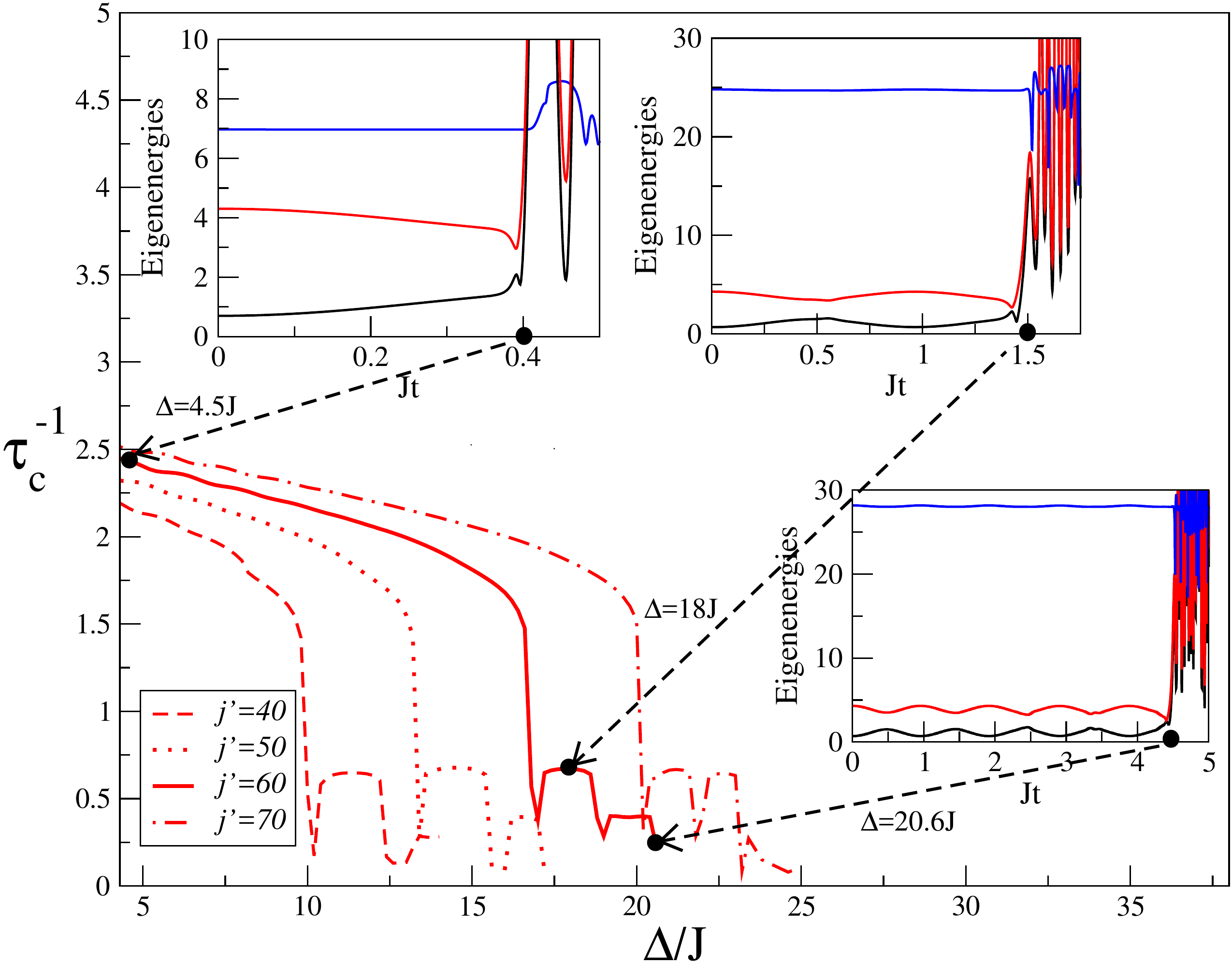}\vspace*{-0.8em} 
\end{center}
\caption{Dependence of the inverse characteristic time $\tau_c^{-1}$ on the
  interlevel spacing $\Delta$ for $z(0)=0.6, \theta(0)=0$ and k=0 for four
  values of the QP-assisted tunneling $j'$. For three values of $\tau_c$ from
  the $j'=60$ curve the time dependence of the energy levels are shown in the 
  three insets as marked. The excitation of QPs is related to the level 
  crossing discussed in the text.}
\label{fig:11}
\end{figure} 
In order to get a better understanding of the reasons of the abrupt QP
generation, we analyzed the instantaneous single-particle levels of the
Hamiltonian \eqref{ham} in Bogoliubov-Hartree-Fock approximation, 
shown in Fig. \ref{fig:9} \cite{Mauro2015}. It turns out that the rapid QP production sets in
when one of the condensate levels (BEC level 2 in Fig. \ref{fig:9}) crosses
or comes close to the first QP level ("QP level 1" in Fig. \ref{fig:9}). In
the case of negligible QP generation (Fig. \ref{fig:8}) the levels
never cross, as seen in the inset of Fig. \ref{fig:9}. 
In view of the afore-mentioned physics it is clear that reducing interlevel
spacing should accelerate QP production. This indeed happens and is
demonstrated in Fig. \ref{fig:10}. 
We reduce $\Delta$ by 25 $\%$, and as a result $\tau_c$ decreases by about 
80 $\%$ compared to Fig. \ref{fig:6}. However, an analytical
parameter dependence of $\tau_c$ is difficult to obtain, since the transition 
to the QP-dominated regime is controlled by highly non-linear processes. 
Thus, a systematic numerical study of the inverse characteristic time $\tau_c^{-1}$ 
versus $\Delta$ for different  $j'$-s and fixed initial conditions is
presented in Fig. \ref{fig:11}. As expected, $\tau_c^{-1}$ generally
decreases with increasing $\Delta$, but not in a monotonic way. 
Namely, one can distinguish two regimes of qualitatively different behavior of
$1/\tau_c$, separated by the oscillation period $T_{12}$ of the condensate levels
$\alpha=1,2$ for $t<\tau_c$:
For $1/\tau_c>1/T_{12}$, $\tau_c$ depends on $\Delta$ in a continuous way, while 
for $1/\tau_c<1/T_{12}$ it jumps between certain discrete or plateau 
values. This happens because as long as $\tau_c$ is smaller than $T_{12}$,  
the condensates cannot perform a full Josephson oscillation before QPs get
excited. As a result the BEC dynamics cannot be considered a periodic
perturbation on the QP system, and any value of $\tau_c$ is possible, thus
increasing continuously with $\Delta$. 
For larger $\Delta$, $\tau_c$ becomes larger than $T_{12}$, and $\tau_c$ takes on
prefered plateau values which are related to the times when the condensate
levels cross or come very close to each other and the first QP level 
(see insets of Fig. \ref{fig:11}). The detailed discussion of this physics 
and the dependence of $\tau_c$ on the parameter $k$ can be found in \cite{Mauro2015}.

\begin{figure}[t]
\begin{center}
\includegraphics[width=0.7\linewidth]{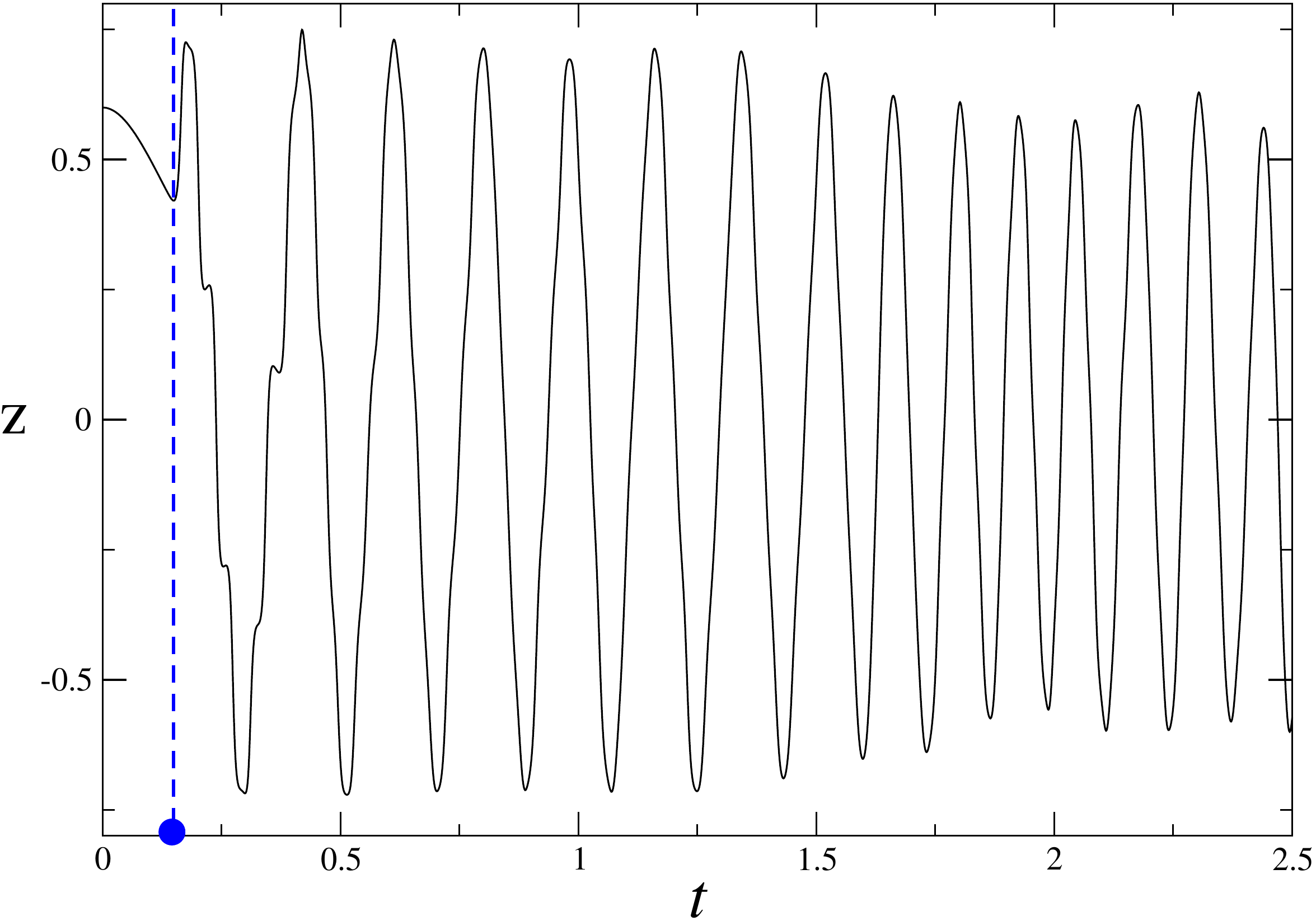}\vspace*{-0.8em} 
\end{center}
\caption{Time-dependent population imbalance in the initially self-trapped case $z(0)=0.6$, $\theta(0)=0$, $\Delta=20, u=u'=22, j'=60, k=10$. The characteristic time scale $\tau_c$ is marked by the dashed vertical line, and $\tau_c$ is shown by the thick dot.}
\label{fig:12}
\end{figure}

Finally, we comment on the self-trapped case. This regime of macroscopic
self-trapping (ST) with a finite time average $\langle z(t) \rangle\neq 0$ and
an unbounded phase difference $\theta(t)$ was predicted in
Ref. \cite{Smerzi1997} and verified experimentally in
Ref. \cite{Albiez2005}. In the preceding discussion we considered the
initially delocalized regime with $\langle z(t)\rangle=0$ and oscillating
$\theta(t)$. For the ST case the non-equilibrium dynamics is very similar to the
delocalized case \cite{Mauro2015}. However, the values of $\Delta$ for which the
QP creation time $\tau_c$ approaches zero are substantially greater
due to the substantially greater BEC oscillation frequencies. It means that 
in the ST case the system is much easier to drive into the QP-dominated regime. 
The initial ST is always destroyed by the QP dynamics, see Fig. \ref{fig:12}. 
For the initially self-trapped case we show the population imbalance dynamics 
only, because all other results are very similar to the delocalized 
case \cite{Mauro2015}. We note also, that first principle calculations on level energies and coupling constants,
based on trap wave functions for realistic traps of the experiments of 
Ref. [34] and [36] are in preparation and will shed light on the precise mechanism of quasi-particle creation.

\subsection{Thermalization by Quasiparticle Collisions}
\label{subsec:thermalization}

We demonstrate how the physics discussed in the previous section \ref{subsec:BHF_results} is
modified by inclusion of QP inelastic collisions (all second-order
processes). We self-consistently solve Eqs. \eqref{spec_fin}, \eqref{stat_fin} and \eqref{cond_fin} (for numerical details see Appendix \ref{numerics}) for initially delocalized junctions.  

\begin{figure}[t]
\begin{center}
\includegraphics[width=0.9\linewidth]{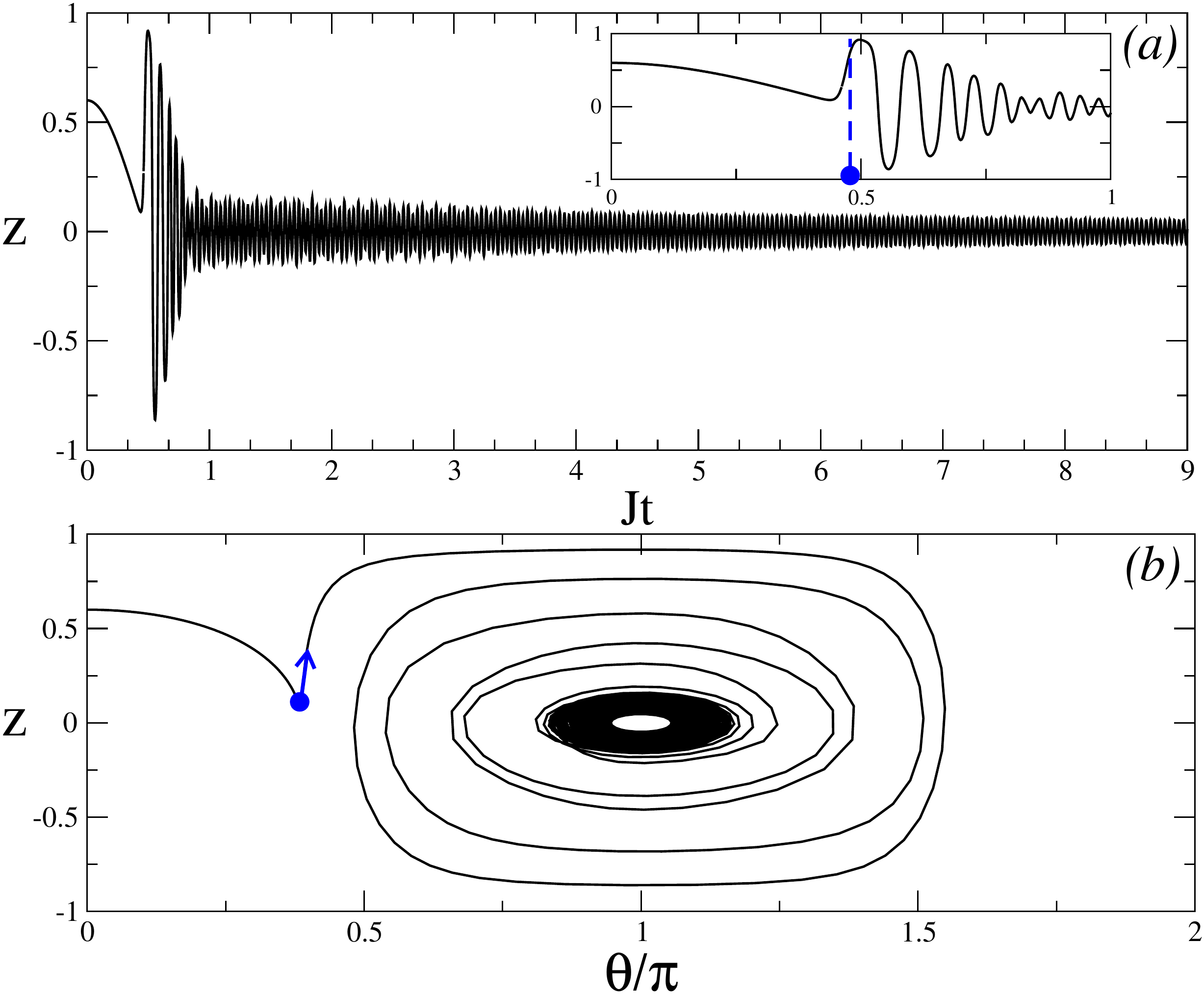}\vspace*{-0.8em} 
\end{center}
\caption{Time-dependent population imbalance $z(t)$ (a) and phase space portrait (b) for $\Delta=10, u=u'=5, j'=60, k=0, r=600$, and initial conditions $z(0)=0.6, \theta(0)=0$. The characteristic time scale $\tau_c$ is marked by the dashed vertical line in the inset, and $\tau_c$ is marked by thick dots. The arrow indicates direction of the time evolution along the phase trajectory after $\tau_c$.}
\label{fig:13}
\end{figure} 
We identify three different regimes and three time scales associated with the 
non-equlibrium dynamics of the Bose Josephson junction (see also Fig. \ref{fig:2}). The regimes are the following.
\begin{enumerate}
\item {\it Semiclassical regime} for $0 \leq t\leq\tau_c$.

QPs are not excited, or their number is negligible, so that the BEC
oscillations are undamped and well described within the two-mode approximation \cite{Smerzi1997}. 

\item {\it Strong coupling regime} for $\tau_c<t<\tau_f$. 

As we know from section \ref{subsec:BHF_results}, the time $\tau_c\geq 0$ marks the onset of
the QP dominated regime. Incoherent excitations are induced in an avalanche
fashion due to a {\it dynamically generated parametric resonance} between the
Josephson frequency and QP excitation energies, as shown below. The BEC and
the QP subsystems are strongly coupled.
This leads to a fast depletion as well as strong damping of the condensate
amplitudes \cite{Anna2016,Yukalov2008}. 

\item  {\it Weak coupling or hydrodynamic regime} for $t>\tau_f$.

At the "freeze-out" time $t=\tau_f>\tau_c$ the final number of excitations
allowed by total energy conservation is reached. This results in an effective
decoupling of the QP subsystem from the BEC oscillations and a near
conservation of the total QP number.  Because of this approximate conservation 
law,  the system enters into a quasi-hydrodynamic regime which is characterized
by exponential relaxation to thermodynamic equilibrium with a slow relaxation
time $\tau_{th}>\tau_f$. The QP subsystem acts as a grand canonical reservoir
for the BEC subsystem and vice versa. Remarkably, we observe that the 
thermalization times $\tau_{th}$ for the BEC and for the QP subsystems may
be different (see below, Fig.~\ref{fig:16}). 
\end{enumerate}

\begin{figure}[t]
\begin{center}
\includegraphics[width=0.96\linewidth]{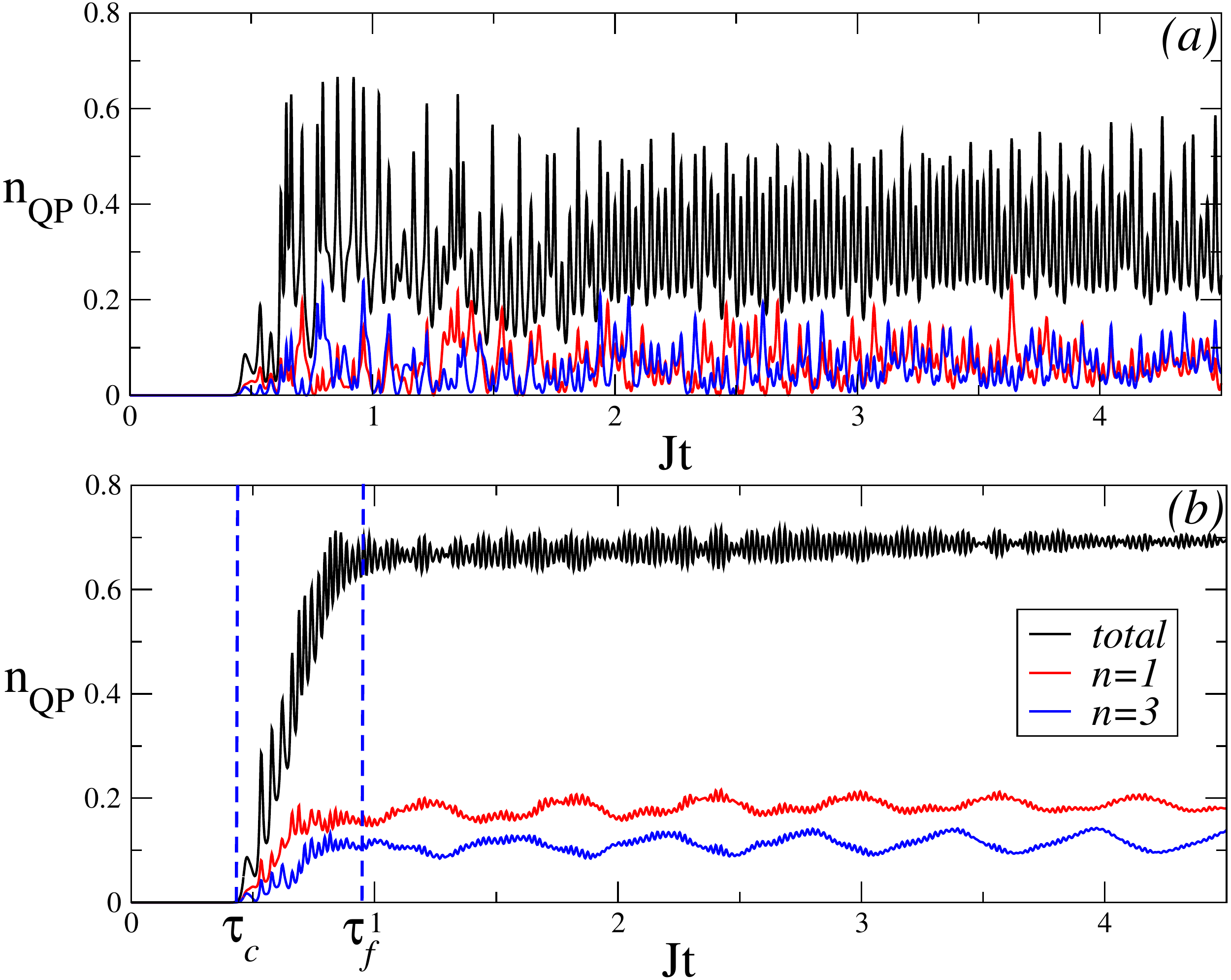}\vspace*{-0.8em} 
\end{center}
\caption{Time-dependent QP occupation numbers for the parameters as in
  Fig. \ref{fig:13} ($\Delta=10, u=u'=5, j'=60, k=0$) in (a)
  Bogoliubov-Hartree-Fock approximation, (b) second-order approximation
  including inelastic collisions ($r=300$). Red lines correspond to the
  occupation of the lowest QP level $n_1$, blue lines to the occupation of the
  third QP level $n_3$, and the black lines represent the occupation numbers
  summed over all 5 levels $n_{\textrm{tot}}$. Dashed vertical lines mark the
  time scales $\tau_c$ and $\tau_f$ discussed at the beginning of 
  section \ref{subsec:thermalization}. }
\label{fig:14}
\end{figure}

We now illustrate this intricate non-equilibrium dynamics with our numerical
results. In Fig. \ref{fig:13} we show an example of the population
imbalance damping, and a phase portrait corresponding to the relaxation.  We
see that the $0-\pi$ transition survives and that the amplitude of the phase
oscillations becomes smaller as one evolves in time, as expected. In
Fig. \ref{fig:14} we compare QP occupation numbers calculated in the
first-order approximation Fig. \ref{fig:14}(a), and in full second-order
Fig. Fig. \ref{fig:14}(b). We see that compared to BHF the oscillations
are strongly damped, although the average QP number can be even greater in the
collisional regime. Three different regimes (semiclassical, strong coupling and weak coupling) are clearly distinguishable in Fig. \ref{fig:14}. 

\begin{figure}[b]
\begin{center}
\includegraphics[width=0.9\linewidth]{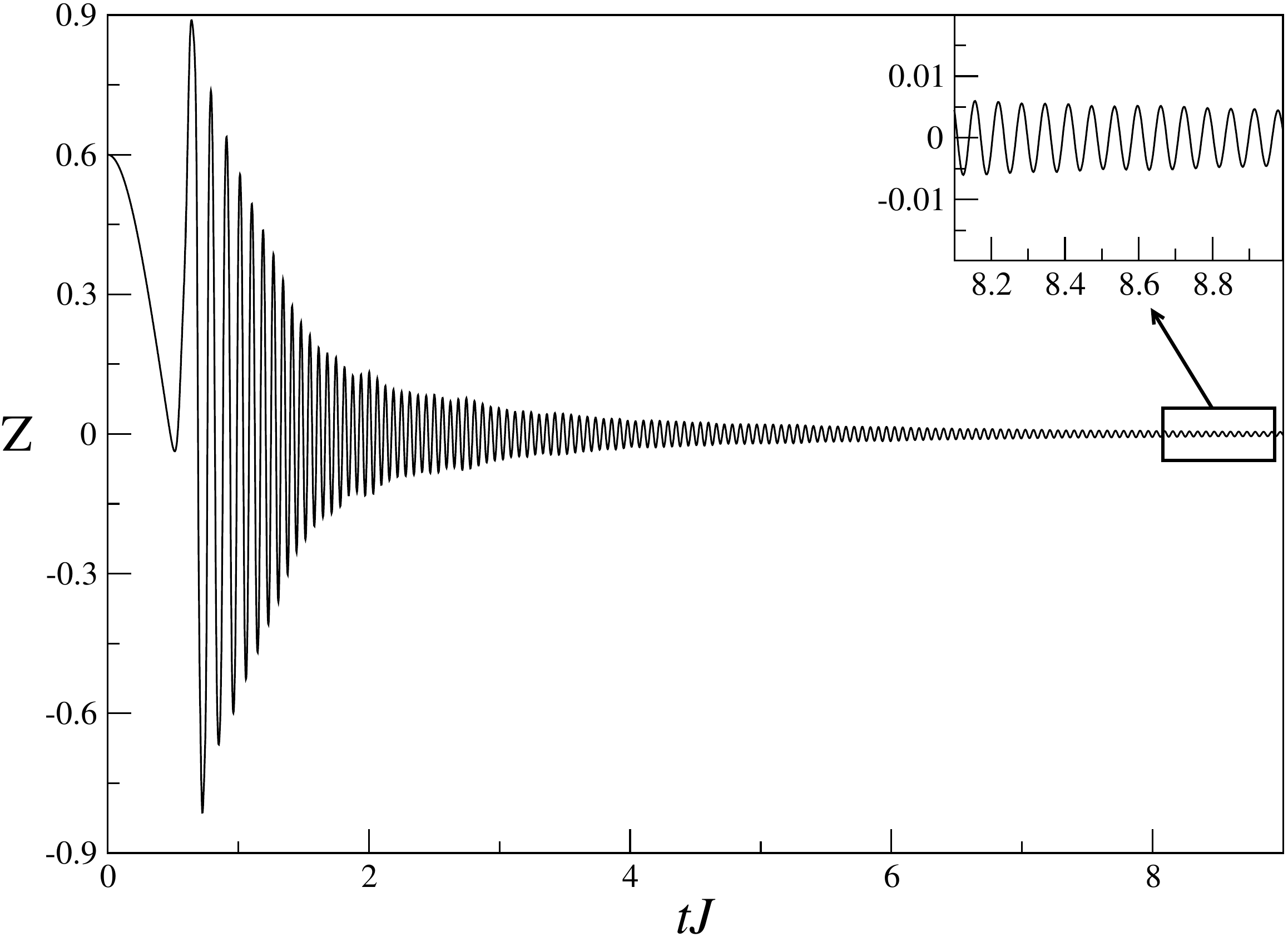}\vspace*{-0.8em} 
\end{center}
\caption{Time-dependent population imbalance  for $z(0)=0.6$, $\theta(0)=0$, $\Delta=9, u=u'=5, j'=40, k=0, r=300$. The inset shows small oscillations remaining in the long time limit. }
\label{fig:15}
\end{figure} 

To understand the origin of this behavior, we now consider a similar junction
in more detail  \cite{Anna2016} (see Fig. \ref{fig:15} for parameter
values). In Fig. \ref{fig:16} we present logarithmic plots of (a) deviation of
the running mean value $n_{\textrm{avg}}(t)$ of $n_{\textrm{tot}}(t)$ from its
final value $n_{\textrm{avg}}(\infty)$; (b) $\Delta
n(t)=n_{\textrm{tot}}(t)-n_{\textrm{avg}}(t)$, and (c) condensate population
imbalance $z(t)$. All three logarithmic plots demonstrate the sharp crossover
at $t=\tau_f$ from the strong to the weak coupling regimes and slow
exponential relaxation for $t>\tau_f$.

\begin{figure}[b]
\begin{center} 
\includegraphics[width=0.9\linewidth]{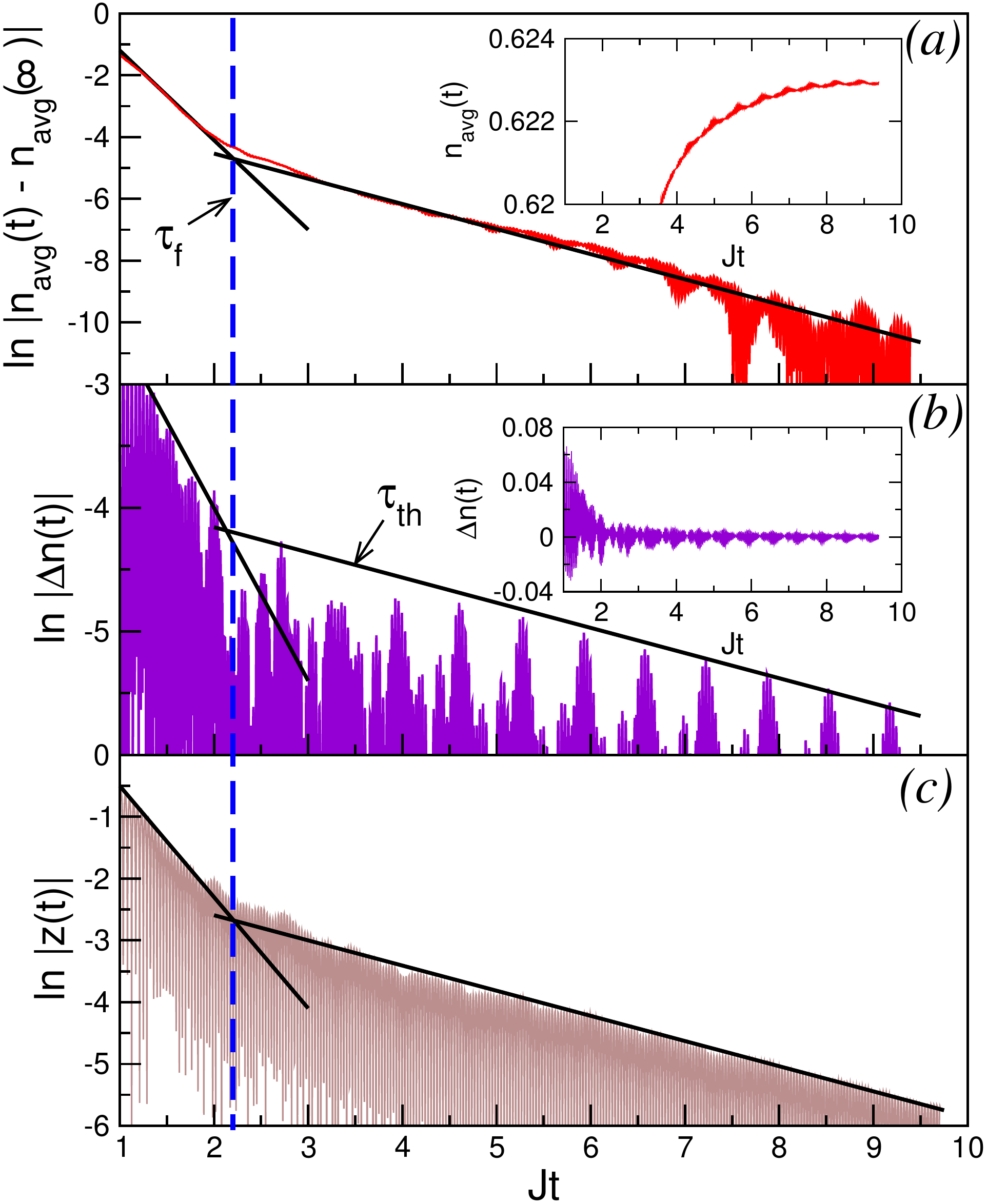}\vspace*{-0.8em} 
\end{center}
\caption{Logarithmic plots of the relaxation behavior of the QP subsystem 
(a) and (b), and of the BEC population imbalance (c) for the same parameters as in Fig. \ref{fig:15}. The dashed vertical line marks the freeze-out time $\tau_f$. 
The thin, black lines are guides to the eye. 
The insets show the respective linear plots, for illustration. \cite{Anna2016} Copyright 2016 by the American Physical Society. }
\label{fig:16}
\end{figure} 

\begin{figure}[t]
\begin{center}
\includegraphics[width=0.96\linewidth]{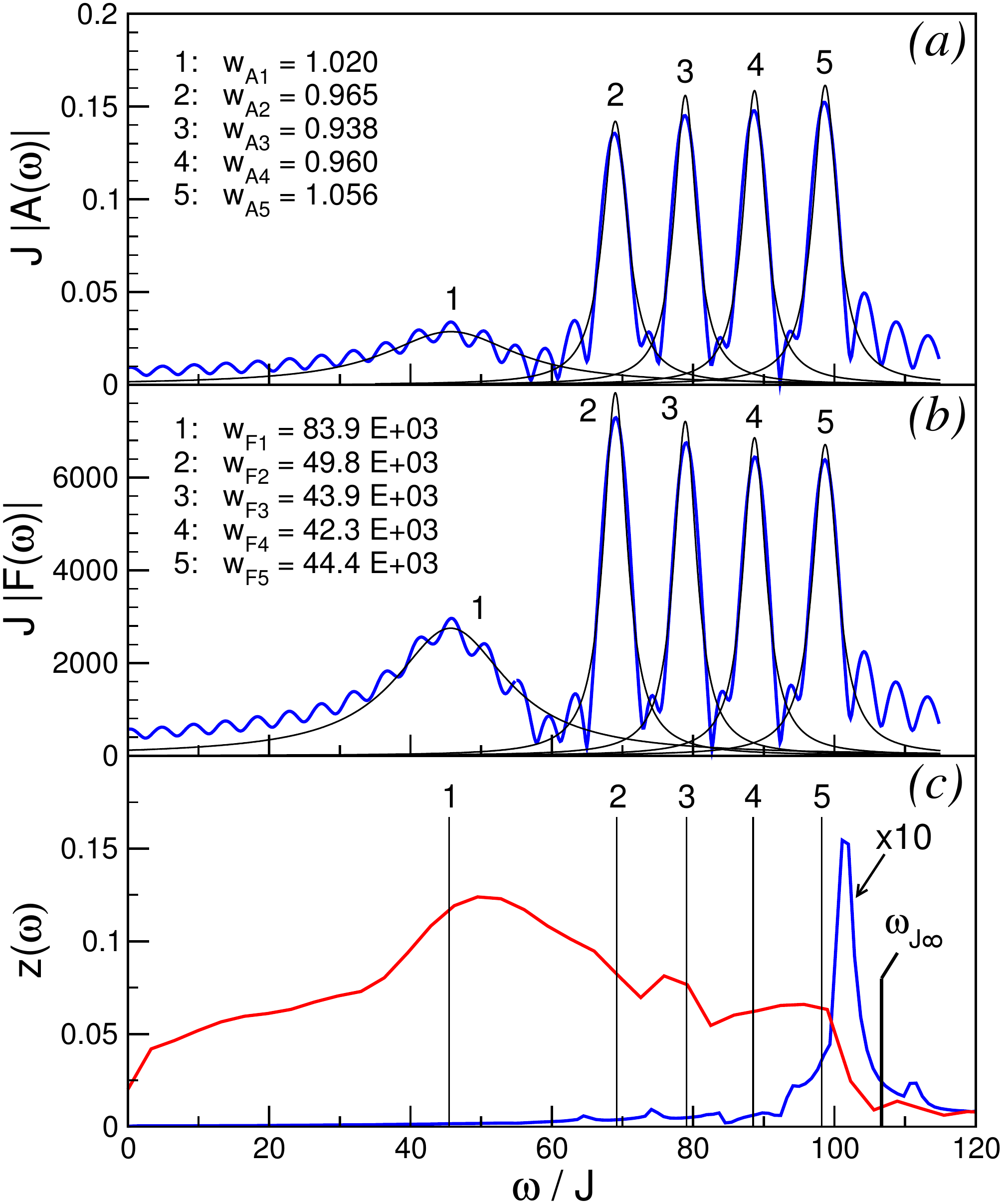}\vspace*{-0.8em} 
\end{center}
\caption{Absolute values of (a) spectral  and (b) statistical  functions, 
  Fourier-transformed with respect to $\tau=(t_1-t_2)$ for a
  fixed value of $t=(t_1+t_2)/2=9.01/J$. The thin, black lines represent 
  Lorentzian fits. The weights $\text{w}$ of each of the five 
  Lorentzians are shown in the insets. In (c) the power 
  spectrum $z(\omega)$ of the BEC population imbalance is shown for 
  $\tau_c\lesssim t \lesssim \tau_f$ (red line) and for $t>\tau_f$ (blue
  line). The vertical lines indicate renormalized QP energies. 
  $\omega_{J\infty}$ is the Josephson frequency estimated for the 
  quasi-hydrodynamic regime for $t>\tau_f$. \cite{Anna2016} Copyright 2016 by the American Physical Society. }
\label{fig:17}
\end{figure}

The physics behind the sharp crossover and the scale $\tau_f$ can be deduced from a spectral analysis of the non-equilibrium problem. We introduce the standard Wigner ``center-of motion'' (CoM) time 
$t=(t_1+t_2)/2$ and difference time $\tau =(t_1 - t_2)$ and Fourier-transform 
the two-time Green's functions 
$\text{A}^G(t_1,t_2)=\sum_n\text{A}_{nn}^G(t_1,t_2)$ 
and $\text{F}^G(t_1,t_2)=\sum_n\text{F}_{nn}^G(t_1,t_2)$ with respect to
$\tau$. 
Note that away from equilibrium the Fourier-transformed functions
are in general complex.
We choose as the zero of the energy scale the renormalized 
energy $\tilde\varepsilon_0$ of the BEC in the long-time limit after
the stationary state has been reached. In particular, this implies  that 
the chemical potential in this final state is $\mu=0$.
In Fig. \ref{fig:17} (a) and (b) we plot the 
frequency-dependent absolute values of
$\text{A}^G(\omega,t)\equiv \text{A}(\omega)$ and
$\text{F}^G(\omega,t)\equiv \text{F}(\omega)$ in the long-time regime,
$t=9.01 /J >\tau_f$.  As expected, the spectra exhibit 
five nearly  Lorentzian peaks corresponding to the 
renormalized QP levels. They mark the Rabi 
oscillation frequencies of the non-equilibrium QP system.
The wiggly modulations of the 
Lorentzian peaks are due to a limited resolution of the Fourier transform \cite{Anna2016}. 
Fig. \ref{fig:17} (c) displays the power spectra of the 
BEC population imbalance $z(t)$, Fourier transformed 
with respect to $t$ for $\tau_c<t<\tau_f$ (red curve) and for $t>\tau_f$ 
(blue curve), respectively. 

The remarkable feature seen in Fig. \ref{fig:17} is that in the strong
coupling regime, the condensate oscillation spectrum overlaps strongly with the
QP spectrum $|\text{A}(\omega)|$ and has maxima approximately at renormalized
Rabi frequencies. This is an indication of a dynamically generated parametric
resonance which leads to an abrupt, "inflationary" QP creation. 
Very different behaviour is observed in the third regime (weak coupling
regime).  The BEC spectrum consists of essentially one sharp 
(compared to the broad spectrum in the strong coupling regime) peak, which has
negligible overlap with the QP spectrum. Moreover, this peak is close to the 
eigenfrequency of the non-driven Josephson junction, which is  
$\omega_{J\infty}\approx 2J_{\textrm{eff}}\sqrt{1+uJ/(2J_{\textrm{eff}})}$,
with $J_{\textrm{eff}}=J+n_{\textrm{tot}}(t\rightarrow \infty)J'$ the
QP-renormalized effective Josephson coupling \cite{Smerzi1997,Anna2016}. 
This manifests that the BEC performs essentially free, 
non-driven Josephson oscillations, i.e., the BEC and the QP subsystems are
effectively decoupled in this final regime. 

The emergence of the weak coupling regime can be understood from energy conservation arguments. The energy of the condensate subsystem can be calculated as the expectation value of the coherent parts of the Hamiltonian 
only, $H_{coh}$ and $H_J$. 
Hence, the general expression for the BEC energy in these regimes is,
\bea
E_{BEC}&=&\sum_{\alpha=1,2}\left[
\varepsilon_0 (N_{\alpha}) +\frac{U}{2} (N_{\alpha}-1)N_{\alpha}\right]
\nn \\
&-&2J\sqrt{N_1N_2} -3J'N_{qp} \sqrt{N_1N_2}
\label{E_BEC1}
\ea  
where  $N_{qp}$ is the particle number in the QP subsystem. 
$N_1$, $N_2$ and $N_{qp}$ can be expressed in terms of the total particle number
$N$, the total condensate 
number $N_c$, and the population imbalance $z$ as
\bea
N_1+N_2=N_c\, , \qquad
N_1-N_2=zN_c\, , \qquad
N_c+N_{qp}=N \ .           \nn 
\ea
Inserting this in Eq.~(\ref{E_BEC1}), 
the BEC energy reads in terms of the reduced interaction constants
$u$, $j'$, and the condensate fraction $f=N_c/N$ as
\bea
E_{BEC}&=& \varepsilon_0 fN + 
\left[
\frac{u}{4}f\left( f-\frac{2}{N}\right) +
\frac{u}{4} z^2 f^2 \right. \\ 
&-& \left. f\left(1-\frac{3}{2}j'(1-f)\right)\sqrt{1-z^2}
\right] NJ \ . \nn
\label{E_BEC2}
\ea
Hence, the initial-state energy at $t=0$, i.e., for $\varepsilon_0=0$, 
$f=1$, and $z=z(0)=z_0$, reads,
\bea
E_{BEC}(0)= 
\frac{u}{4}
\left[ (1+z_0^2- 2/N) - \sqrt{1-z_0^2}
\right] NJ \ .
\label{E_BEC0}
\ea
The final-state energy for $t\to\infty$, where 
$\varepsilon_0=\varepsilon_0(\infty)\neq 0$
(renormalized by QP interactions), $f=f_{\infty}<1$  (finite, but 
decoupled QP population), and $z=0$ (BEC oscillations damped out), reads,
\bea
E_{BEC}(\infty)&=& 
\left[
f_{\infty}\frac{\varepsilon_0(\infty)}{J}
+\frac{u}{4}f_{\infty}\left(f_{\infty}-\frac{2}{N} \right) \right. 
\label{E_BECinfty} \\ 
 &-& \left. f_{\infty} \left( 1+ \frac{3}{2} j' (1-f_{\infty}) \right)
\right] NJ \ . \nn
\ea
The final-state parameters $\varepsilon_0(\infty)$ and $f_{\infty}$ 
can be obtained from the numerical solutions of Eqs. \eqref{spec_fin}, \eqref{stat_fin} and \eqref{cond_fin}.

The energy difference $\Delta E_{BEC}=E_{BEC}(0)-E_{BEC}(\infty) $  is in fact
the maximum energy that can be provided to the QP subsystem by the condensate
subsystem. Therefore, the energy of the QP subsystem $E_{QP}(t)$ initially
increases but eventually saturates once the maximum is reached. This happens
for $t>\tau_f$, and the number of QPs stays approximately constant thereafter.
Our numerical computations show that indeed the maximum is attained at
$t\approx \tau_f$. It means that for $t>\tau_f$ both $n_{\textrm{tot}}(t)$ and
$E_{QP}(t)$ become approximately conserved in the grand canonical sense
(particle and energy exchange between the subsystems are allowed, but
time-averages do not change). 
Under these dynamically generated conservation laws, the system enters into a
quasi-hydrodynamic regime with a slow exponential relaxation toward thermal
equilibrium. Note that in this last regime,
the relaxation times $\tau_{th}$ are different for the BEC oscillations,
$z(t)$, and for the QP relaxation (different y-axis scales on the three
panels in Fig.~\ref{fig:16}).

\begin{figure}[!bt]
\begin{center}
\includegraphics[width=0.48\textwidth]{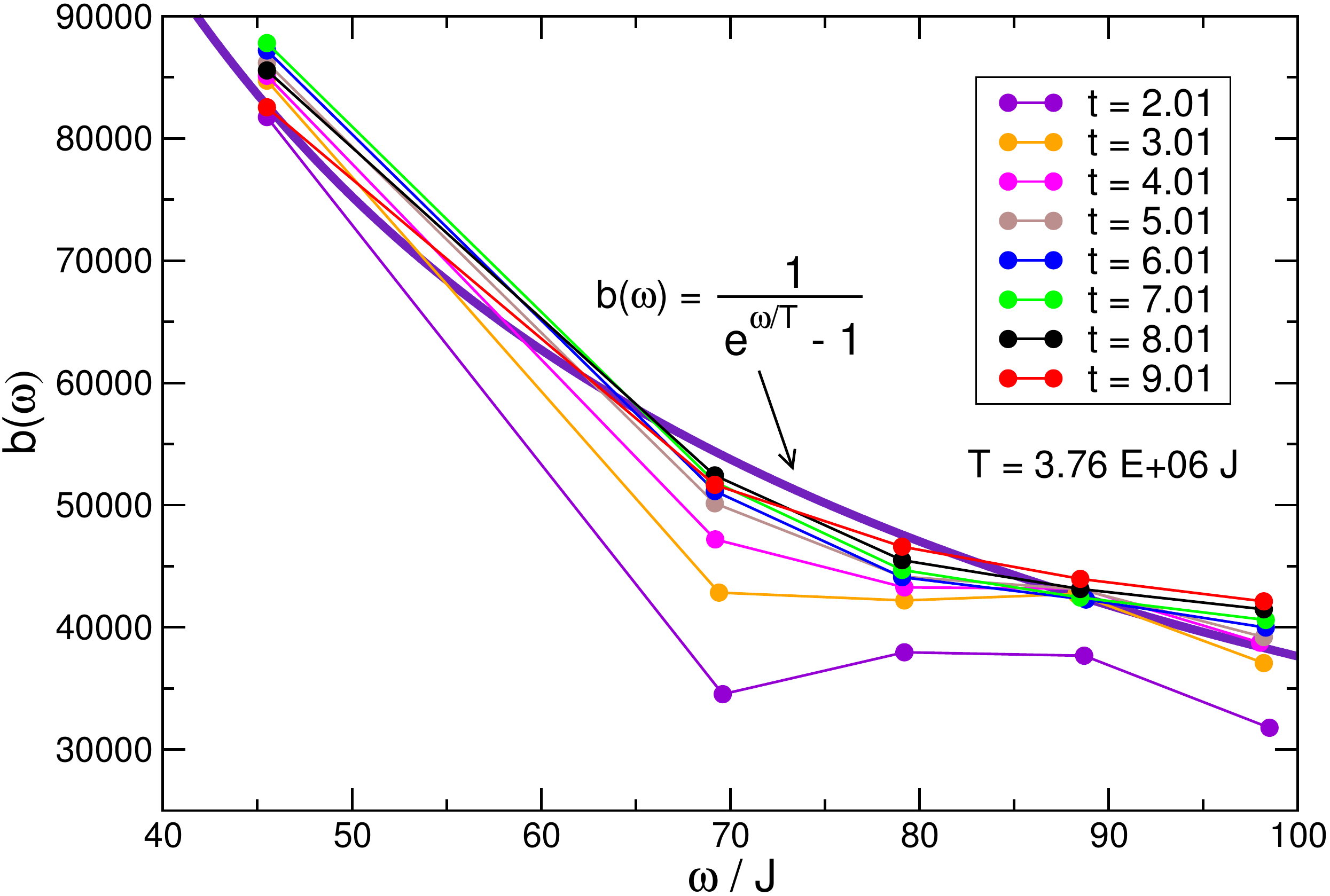}\vspace*{-0.8em} 
\end{center}
\caption{Distribution function $b(\tilde\varepsilon_n,t)$ for different
  CoM times $t$. The thick purple line is a single-parameter
  fit of a thermal distribution to the calculated 
  $b(\tilde\varepsilon_n,t)$  for the largest time $t=9.01$, 
  with temperatue $T$ as fit parameter. The fitted value is 
  $T=3.76\cdot 10^{6}\ J$. \cite{Anna2016} Copyright 2016 by the American Physical Society.}
 \label{fig:18}
 \end{figure} 

To prove that the long-time state is a thermal one, we calculate the QP distribution function $b(\varepsilon_n,t)$
for different CoM times $t$. It is defined via  Keldysh
Green's functions \cite{Rammerbook} by
\beq
F(\omega,t) = (-i/2) (2b(\omega,t)+1) A(\omega,t)
\eq
 and is therefore  
obtained for each level from the Lorentzian weights $w_{A,n}$,
$w_{F,n}$ of these levels (c.f. Fig.~\ref{fig:17}) as
\bea
b(\tilde\varepsilon_n,t)=\frac{w_{F,n}}{w_{A,n}}-\frac{1}{2}  \ .
\ea  
Here $\tilde\varepsilon_n$, $n=1,\, \dots,\, M$, $(M=5)$, are the 
level energies, renormalized by interactions. 
As shown in Fig.~\ref{fig:18},
$b(\tilde\varepsilon_n,t)$ continuously approaches a thermal distribution.
As expected, the final-state temperature $T$ 
is high, since it is controlled by the initial BEC excitation energy,
$\Delta E_{BEC}\sim z(0)^2N_{\textrm{tot}}J$, which is a macroscopically large 
quantity.

\section{Conclusions and Discussion}
\label{sec:conclusions}

We demonstrated that the system of coupled, oscillating BECs and 
incoherent excitations 
thermalizes, because the condensates serve as a heat reservoir for the 
QP subsystem and visa versa. The QP subsystem is generated "naturally" as a result of complex non-equilibrium dynamics, in fact a parametric resonance. At a later time $\tau_f$
 the energy of QP subsystem reaches its maximum value determined by the difference between the condensate initial and final energies, and the two subsystems become essentially decoupled in the grand canonical sense. The main reason for such a decoupling is total energy conservation and entropy maximization in the 
QP subsystem  \cite{Anna2016}.  

For times smaller than $\tau_f$, the condensate and the QPs are strongly
coupled which is clearly  seen in the resonating spectra of the two subsystems. 
For times $t>\tau_f$,  BEC and incoherent excitations exhibit off-resonant 
behavior, confirming the decoupling. This is the essence of DBG. 

In the off-resonant regime, the QP system
relaxes slowly to a high-temperature thermal state  with thermalization time 
$\tau_{th} > \tau_f$. The BEC freeze-out and subsequent time evolution under a
conservation law are reminiscent of pre-thermalization found in
low-dimensional, nearly integrable systems \cite{Joerg_review}. 
However, our system is non-integrable, and the (approximate) 
conservation law is dynamically generated. 

Remarkably, the non-equilibrium dynamics of the trapped Bose-gas system 
resembles the preheating and thermalization dynamics of 
inflationary models of the early universe \cite{Kofman1994}, see also 
Ref.~\cite{Zache2017}. The Bose gas is initially prepared in a 
non-equilibrium state, analogous to the inflationary period of the 
early universe. The coherent Josephson oscillations of the BEC subsystem
correspond to the the excitations of an inflaton field, postulated by the 
early-universe models \cite{Kofman1994}. The creation of QP excitation in 
the Bose gas system represents the creation of elementary particles 
after the inflationary period of the early universe. 
In both, the Bose gas system and the early universe, a parametric resonance 
emerges dynamically, between BEC Josephson oscillations and QP excitations 
on one hand, and between inflaton-field oscillations and elementary 
particles on the other hand.  Finally, the effective
decoupling of our bosonic subsystems corresponds to the inflaton decoupling 
due to loss of resonance by expansion of the universe. 
These analogies are worth further exploration \cite{Zache2017}.

\acknowledgments
We gratefully acknowledge fruitful discussions with  
Stefan Kehrein, Tim Lappe, Marvin Lenk, and J\"org Schmiedmayer. 
This work was supported in part by the Deutsche 
Forschungsgemeinschaft (DFG) through SFB/TR 185 (J.K).


\appendix


\section{Notes on Numerical Implementation}

\label{numerics}

\begin{figure}[t]
\begin{center}
\includegraphics[width=\linewidth]{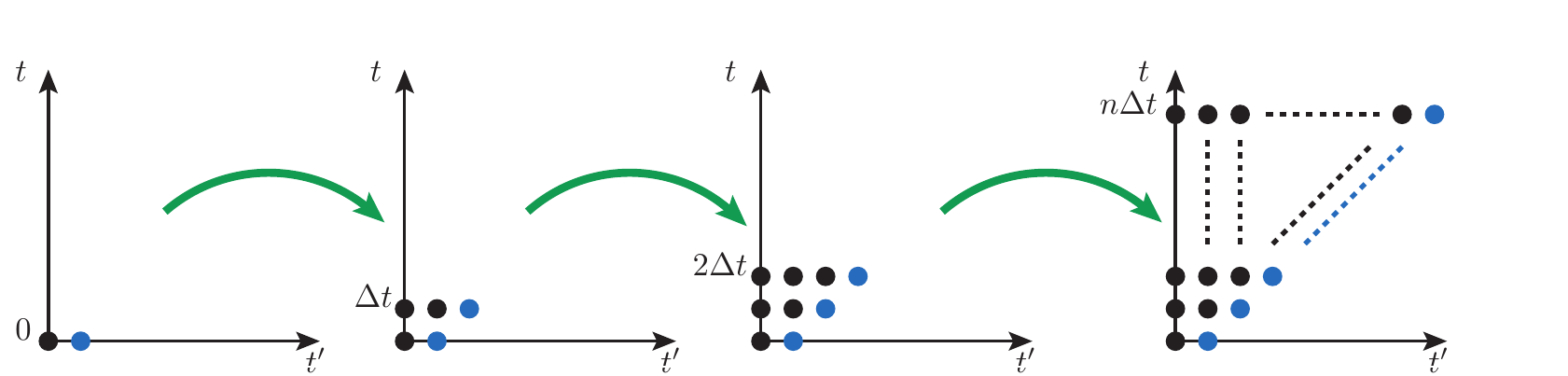}\vspace*{-0.8em} 
\end{center}
\caption{Evolution of the time grid in the two-time plane.}
\label{fig:19}
\end{figure}

In order to solve numerically our system of integro-differential equations, we discretize the two time arguments, $t$ and $t'$ with a constant time-step $\Delta t$ (see Fig. \ref{fig:19}). As a result, our spectral and statistical functions become matrices in the two-dimensional time plane. For instance, 
\bea
\text{F}(t,t')=\begin{pmatrix}
\text{F}(0,0) & \text{F}(0,\Delta t) & \cdots & \text{F}(0,n\Delta t) \\
\text{F}(\Delta t,0) & \text{F}(\Delta t,\Delta t) & \cdots & \text{F}(\Delta t, n\Delta t) \\
\vdots & \vdots & \ddots & \vdots \\
\text{F}(n\Delta t,0) &\text{F}(n\Delta t, \Delta t) & \cdots & \text{F}(n\Delta t,n\Delta t)
\end{pmatrix} \nn \\
\label{full_matrix}
\ea
where both time arguments are counted from $\tau_c$, which is a finite time-scale at which the non-equilibrium dynamics sets in \cite{Mauro2015}, so that $\text{F}^G(0,0)\equiv \text{F}^G(\tau_c,\tau_c)$ etc. Both time scales go up to $t_{max}=n\Delta t$. In our case $t_{max}=10$. Fortunately, due to the symmetry relations \eqref{symmetry} it is sufficient to calculate only half of the components of our propagators, i.e. the triangular matrix
\bea
\text{F}(t,t')_{tri}=\begin{pmatrix}
\text{F}(0,0) & \text{F}(0,\Delta t) & \cdots & \text{F}(0,n\Delta t) \\
0 & \text{F}(\Delta t,\Delta t) & \cdots & \text{F}(\Delta t, n\Delta t) \\
\vdots & \vdots & \ddots & \vdots \\
0 & 0 & \cdots & \text{F}(n\Delta t,n\Delta t),
\end{pmatrix} \nn \\
\ea
which is reflected in the time-plane grid in Fig.\ref{fig:19}. The blue points on the grid constitute additional copies of the diagonal (propagators with equal time arguments) contributions, necessary to properly perform the fourth order Runge-Kutta method. 
Symmetry relations for the self-energies \eqref{symmetry_self} also contribute to simplifications, as we can rewrite all the integrands in Eqs. \eqref{spec_fin}, \eqref{stat_fin} with time argument corresponding to later time on the left, e.g. time convolutions in the equation of motion for $F^G$ can be rewritten as
\bea
-i\int_{0}^t d{\ov t}\left[\Gamma^G_{n\ell}(t,{\ov t})\text{F}^G_{\ell m}({\ov t,t'})+\Gamma^F_{n\ell}(t,{\ov t})\text{F}^{\ov F}_{\ell m}({\ov t},t')\right] \nn \\
+i\int_{0}^t d{\ov t}\left[\Pi^G_{n\ell}(t,{\ov t})\text{A}^G_{\ell m}({\ov t,t'})+\Pi^F_{n\ell}(t,{\ov t})\text{A}^{\ov F}_{\ell m}({\ov t},t')\right]  \nn\\
=i\int_{0}^{t'} d{\ov t}\left[\Gamma^G_{n\ell}(t,{\ov t})\text{F}^G_{\ell m}(t',{\ov t})^*+\Gamma^F_{n\ell}(t,{\ov t})\text{F}^{F}_{\ell m}(t',{\ov t})^*\right] \nn \\
-i\int_{t'}^t d{\ov t}\left[\Gamma^G_{n\ell}(t,{\ov t})\text{F}^G_{\ell m}({\ov t,t'})-\Gamma^F_{n\ell}(t,{\ov t})\text{F}^{F}_{\ell m}({\ov t},t')^*\right] \nn \\
+i\int_{0}^{t'} d{\ov t}\left[\Pi^G_{n\ell}(t,{\ov t})\text{A}^G_{\ell m}(t',{\ov t})^*+\Pi^F_{n\ell}(t,{\ov t})\text{A}^{F}_{\ell m}(t',{\ov t})^*\right]  \nn \\
\ea
In non-equilibrium it is conventional to introduce mixed or Wigner coordinates: $\tau=t-t'$ and $T=(t+t')/2$ and then Fourier transform spectral functions and statistical functions with respect to the relative coordinate. In this way one can extract information about spectrum and distribution function for different values of $T$ and check if system approaches equilibrium with increasing $T$. In our case it is done by reading off the calculated spectral and statistical functions belonging to diagonals with the slope equal to $-1$ from the $t-t'$ plane in Fig. \ref{fig:19}. Those will be data for fixed $T$-s. We then Fourier transform them with respect to $\tau$. We checked the numerical accuracy by varying the time step $dt$ used 
in the differential equation solver. All the results are 
reproducible and independent of $dt$.

\section{Convolution Integrals in Equations of Motion}
\label{integrals}

We  use the symmetry relations  \eqref{symmetry}, \eqref{symmetry_gamma}, \eqref{symmetry_self} also in the convolution integrals which enter our equations of motion \eqref{spec_fin} and \eqref{stat_fin}. Consider, for example, integrals in  \eqref{stat_fin}
\begin{equation}
\begin{aligned}
&-\mathbbm{i}\sum_k\int \limits_{0}^{t}d \ov{t}[\Gamma^G_{ik}(t,\ov{t})\text{F}^G_{kj}(\ov{t},t')+\Gamma^F_{ik}(t,\ov{t})\text{F}^{\ov{F}}_{kj}(\ov{t},t')]  \\
&+\mathbbm{i}\sum_k\int \limits_{0}^{t'}d \ov{t}[\Pi^G_{ik}(t,\ov{t})\text{A}^G_{kj}(\ov{t},t')+\Pi^F_{ik}(t,\ov{t})\text{A}^{\ov{F}}_{kj}(\ov{t},t')]. 
\end{aligned}
\label{int_1}
\end{equation}
The symmetry relations allow us to split the interval of integration in such a way that we can rewrite the integrals with the arguments corresponding to the later time as first arguments. Hence we get for integral \eqref{int_1} 
\begin{equation}
\begin{aligned}
&\mathbbm{i}\sum_k\int \limits_{0}^{t'}d \ov{t}[\Gamma^G_{ik}(t,\ov{t})\text{F}^G_{jk}(t',\ov{t})^*+\Gamma^F_{ik}(t,\ov{t})\text{F}^{F}_{jk}(t',\ov{t})^*]  \\
&-\mathbbm{i}\sum_k\int \limits_{t'}^{t}d \ov{t}[\Gamma^G_{ik}(t,\ov{t})\text{F}^G_{kj}(\ov{t},t')-\Gamma^F_{ik}(t,\ov{t})\text{F}^{F}_{kj}(\ov{t},t')^*] \\
&+\mathbbm{i}\sum_k\int \limits_{0}^{t'}d \ov{t}[\Pi^G_{ik}(t,\ov{t})\text{A}^G_{jk}(t',\ov{t})^*+\Pi^F_{ik}(t,\ov{t})\text{A}^{F}_{jk}(t',\ov{t})]. 
\end{aligned}
\end{equation}
With the other integrals of Eqs. \eqref{spec_fin},\eqref{stat_fin},
we proceed in analogous way and then solve the final system of equations 
numerically.


\begin{thebibliography}{99}

\bibitem{Trotzky2012}
S. Trotzky, Y.-A. Chen, A. Flesch, I. P. McCulloch, U. Schollw\"ock,
J. Eisert,  I. Bloch, Nature Phys. {\bf 8}, 325 (2012). 

\bibitem{Gring2012} M. Gring, M. Kuhnert, T. Langen, T. Kitagawa, B. Rauer,
  M. Schreitl, I. Mazets, D. A. Smith, E. Demler,  J. Schmiedmayer,
  Science {\bf 337}, 1318 (2012).

\bibitem{Polkovnikov2011} A. Polkovnikov, K. Sengupta, A. Silva,  M. Vengalattore, Rev. Mod. Phys. {\bf 83}, 863 (2011).

\bibitem{Yukalov2011} V. I. Yukalov, Laser Phys. Lett. {\bf 8}, 485 (2011).

\bibitem{Deutsch1991} 
J. M. Deutsch, Phys. Rev. A {\bf 43}, 2046 (1991).

\bibitem{Srednicki1994}
M. Srednicki, Phys. Rev. E {\bf 50}, 888 (1994).

\bibitem{Rigol2008} M. Rigol, V. Danjko,  M. Olshanii, Nature {\bf 452}, 854 (2008). 

\bibitem{ETH_review} L. D'Alessio, Y. Kafri, A. Polkovnikov,  M. Rigol, Advances in Physics {\bf 65}, 239 (2016). 

\bibitem{Goldstein2006} S. Goldstein, J. L. Lebowitz,  R. Tumulka,  N. Zanghi, Phys. Rev. Lett. {\bf 96}, 050403 (2006). 

\bibitem{Reimann2015} P. Reimann, 
Phys. Rev. Lett. {\bf 115}, 010403 (2015). 

\bibitem{Pozsgay2014} B. Pozsgay, J. of Stat. Mech.: Theory and Experiment, P09026 (2014). 

\bibitem{Alba2015} V. Alba, Phys. Rev. B {\bf 91}, 155123 (2015). 

\bibitem{Kollath2007}
C. Kollath, A. L\"auchli,  E. Altmann, Phys. Rev. Lett. {\bf 98},
180601 (2007).

\bibitem{Kehrein2008}
M. Moeckel, S. Kehrein, Phys. Rev. Lett. {\bf 100}, 175702 (2008).

\bibitem{Kollar2011}
M. Kollar,  F. A. Wolf,  M. Eckstein,
Phys. Rev. B {\bf 84}, 054304 (2011).

\bibitem{Joerg_review} T. Langen, T. Gansenzer,  J. Schmiedmayer, {\it Prethermalization and universal dynamics in near-integrable quantum systems}, cond-mat arXiv:1603.09385 (2016). 

\bibitem{Mauro2009} M. Trujillo-Martinez, A. Posazhennikova,  J. Kroha, Phys. Rev. Lett. {\bf 103}, 105302 (2009). 

\bibitem{Mauro2015} M. Trujillo-Martinez, A. Posazhennikova,  J. Kroha, New. J. Phys. {\bf 17}, 013006 (2015). 

\bibitem{Anna2016} A. Posazhennikova, M. Trujillo-Martinez,  J. Kroha, Phys. Rev. Lett. {\bf 116}, 225304 (2016). 

\bibitem{Landau} L. D. Landau, E. M.  Lifshits, {\it Statistical Physics}, Volume V, Elsevier (1980). 

\bibitem{Khinchin} A. I. Khinchin, {\it Mathematical foundations of statistical mechanics}, Dover (1960).

\bibitem{Sinai} Ya. G. Sinai, {\it Introduction to Ergodic Theory}, Princeton University Press (1977). 

\bibitem{Birkhoff} G. D. Birkhoff, Proc. Natl. Acad. Sci USA {\bf 17}, 656 (1931).

\bibitem{Neumann} J. von Neumann, Zeitschrift f\"ur Physik {\bf 57}, 30 (1929).

\bibitem{Singh} N. Singh, Mod. Phys. Lett. B {\bf 27}, 1330003 (2013). 

\bibitem{Rigol2012} M. Rigol, M. Srednicki, Phys. Rev. Lett. {\bf 108}, 110601 (2012). 

\bibitem{Werner2014} H. Aoki, N. Tsuji, M. Eckstein, M. Kollar, T. Oka,  P. Werner, Rev. Mod. Phys. {\bf 86}, 779 (2014). 

\bibitem{Werner2015} H. U. R. Strand, M. Eckstein,  P. Werner, Phys. Rev. X {\bf 5}, 011038 (2015). 

\bibitem{Schollwoeck2005} U. Schollw\"ock, Rev. Mod. Phys. {\bf 77}, 259 (2005). 
\bibitem{Roux2009} G. Roux, Phys. Rev. A {\bf 79}, 021608(R) (2009).

\bibitem{Josephson1962} B. D. Josephson, Phys. Lett. {\bf 1}, 251 (1962). 

\bibitem{Javanainen1986} J. Javanainen, Phys. Rev. Lett. {\bf 57}, 3164 (1986). 

\bibitem{Smerzi1997} 
A. Smerzi, S. Fantoni, S. Giovanazzi,  S. R. Shenoy, 
Phys. Rev. Lett. {\bf 79}, 4950 (1997).

\bibitem{Albiez2005} 
M. Albiez, R. Gati, J. F\"olling, S. Hunsmann, M. Cristiani, 
M. K. Oberthaler, Phys. Rev. Lett. {\bf 95}, 010402 (2005). 

\bibitem{Levy2007} S. Levy, E. Lahoud, I. Shomroni, J. Steinheuer, 
Nature {\bf 449}, 579 (2007). 

\bibitem{Thywissen2011} L. J. LeBlanc,  A. B. Bardon, J. McKeever, 
M. H. T. Extavour, D. Jervis, J. H. Thywissen, F. Piazza,  A. Smerzi, 
Phys. Rev. Lett. {\bf 106}, 025302 (2011).

\bibitem{Milburn1997}  G. Milburn, J. Corney, E. Wright, D. Walls,  Phys. Rev. A {\bf 55}, 4318 (1997)

\bibitem{Gati2007} R. Gati, M. K. Oberthaler, J. Phys. B: At. Mol. Opt. Phys. {\bf 40}, R61 (2007). 

\bibitem{Zapata98} I. Zapata, F. Sols,  A. J. Leggett, 
Phys. Rev. A {\bf 57}, R28 (1998). 

\bibitem{Zapata03}  I. Zapata, F. Sols,  A.\,J. Leggett,
Phys. Rev. A {\bf 67}, 021603(R) (2003).

\bibitem{Pitaevskii01} L. Pitaevskii, S. Stringari, Phys. Rev. Lett. {\bf 87}, 180402 (2001).

\bibitem{Esteve08} J. Esteve {\it et al.}, Nature (London) {\bf 455}, 1216 (2008).

\bibitem{Smerzi2003} A. Smerzi, A. Trombettoni, Phys. Rev. A {\bf 68}, 023613 (2003).

\bibitem{Gati_diss} R. Gati, {\it Bose-Einstein Condensates in a Single Double Well Potential}, Dissertation, 2007. 

\bibitem{Ananikian2006} D. Ananikian, T. Bergeman, Phys. Rev. A {\bf 73}, 013604 (2006). 

\bibitem{Berges_review} J. Berges, "Introduction to Nonequilibrium Quantum Field Theory", AIP Conf. Proc. {\bf 739}, 3 (2004).

\bibitem{Rey2005} A. M. Rey, B. L. Hu, E. Calzetta,  C. W. Clark, Phys. Rev. A {\bf 72}, 023604 (2005). 

\bibitem{Rammerbook} J. Rammer, 
{\it Quantum Field Theory of Non-equilibrium States}, 
Cambridge University Press (2007).

\bibitem{Kadanoffbook} L. P. Kadanoff,  G. Baym, {\it Quantum Statistical Mechanics}, New York: Banjamin (1968). 

\bibitem{Griffinbook} A. Griffin, T. Nikuni,  E. Zaremba, {\it Bose-Condensed Gases at Finite Temperatures}, Cambridge University Press (2009). 

\bibitem{Mauro2017} M. Trujillo-Martinez, A. Posazhennikova,  J. Kroha, in preparation. 

\bibitem{Lappe2017} T. Lappe, A. Posazhennikova,  J. Kroha, in preparation. 

\bibitem{Yukalov2008} V. I. Yukalov,  E. P. Yukalova, Phys. Rev. A {\bf 78}, 063610 (2008). 

\bibitem{Kofman1994} L. Kofman, A. Linde,  A. A. Starobinsky, Phys. Rev. Lett. {\bf 73}, 3195 (1994). 

\bibitem{Zache2017}
T. V. Zache, V. Kasper,  J. Berges, arXiv:1704.02271 (2017).


\end{thebibliography}
\end{document}